\documentclass[amsmath,superscriptaddress,twocolumn]{revtex4-2} 

\usepackage[utf8]{inputenc}
\usepackage{amssymb}
\usepackage{amsmath}
\usepackage{amsthm}
\usepackage{amsfonts}
\usepackage[colorlinks=true,linkcolor=teal,citecolor=teal,urlcolor=teal]{hyperref}
\usepackage{bbold}
\usepackage{bm}
\usepackage{bbm}
\usepackage{microtype}
\usepackage{graphicx}
\usepackage{xcolor}

\usepackage{comment}
\usepackage{times}
\usepackage{epsfig}

\usepackage{listings}
\usepackage{soul}
\usepackage{multirow}
\usepackage{algorithm}
\usepackage[capitalise]{cleveref}

\usepackage{pdfpages}
\usepackage{pgffor}
\makeatletter
\AtBeginDocument{\let\LS@rot\@undefined}
\makeatother

\newtheorem{lemma}{Lemma}
\newtheorem{theorem}{Theorem}

\newtheorem{definition}{Definition}

\newcommand{\norm}[2][]{
	\ifthenelse{\equal{#1}{}}
	{\left\| {#2} \right\|}
	{\ifthenelse{\equal{#1}{uinv}}
		{\left\vert\kern-0.25ex\left\vert\kern-0.25ex\left\vert {#2} \right\vert\kern-0.25ex\right\vert\kern-0.25ex\right\vert}
		{\left\| {#2} \right\|_{#1}}
	}
}

\newcommand{\taverage}[2][]{
	\ifthenelse{\equal{#1}{}}
	{\overline{#2}}
	{\overline{#2}^{#1}}
}

\newcommand{\ee}{\mathrm{e}}
\newcommand{\ii}{\mathrm{i}}

\newcommand{\tracedistance}[3][]{
	\ifthenelse{\equal{#2}{}}
	{\ifthenelse{\equal{#3}{}}
		{\mathcal{D}_{#1}}{}
	}{
		\ifthenelse{\equal{#1}{}}
		{\mathchoice{\operatorname{\mathcal{D}}\left(#2,#3\right)}{\operatorname{\mathcal{D}}(#2,#3)}{\operatorname{\mathcal{D}}(#2,#3)}{\operatorname{\mathcal{D}}(#2,#3)}}
		{\mathchoice{\operatorname{\mathcal{D}}_{#1}\left(#2,#3\right)}{\operatorname{\mathcal{D}}_{#1}(#2,#3)}{\operatorname{\mathcal{D}}_{#1}(#2,#3)}{\operatorname{\mathcal{D}}_{#1}(#2,#3)}}
	}
}
\newcommand{\fidelity}[3][]{
	\ifthenelse{\equal{#2}{}}
	{\ifthenelse{\equal{#3}{}}
		{\mathcal{F}_{#1}}{}
	}{
		\ifthenelse{\equal{#1}{}}
		{\mathchoice{\operatorname{\mathcal{F}}\left(#2,#3\right)}{\operatorname{\mathcal{F}}(#2,#3)}{\operatorname{\mathcal{F}}(#2,#3)}{\operatorname{\mathcal{F}}(#2,#3)}}
		{\mathchoice{\operatorname{\mathcal{F}}_{#1}\left(#2,#3\right)}{\operatorname{\mathcal{F}}_{#1}(#2,#3)}{\operatorname{\mathcal{F}}_{#1}(#2,#3)}{\operatorname{\mathcal{F}}_{#1}(#2,#3)}}
	}
}

\newcommand{\Sr}[3][]{
	\ifthenelse{\equal{#1}{}}
	{\operatorname{\mathnormal{S}}(#2\|#3)}
	{\operatorname{\mathnormal{S}}_{#1}(#2\|#3)}
}



\newcommand{\tr}[1]{\mathrm{tr}\left(#1\right)}

\providecommand{\bgreek}[1]{\mbox{\boldmath$#1$}}

\definecolor{jens}{rgb}{0,0,0}

\newcommand{\newtext}[1]{{\color{blue} #1}}
\begin{document}
	


\title{Typical thermalization of low-entanglement states}

	\author{Christian Bertoni}

	\affiliation{Dahlem Center for Complex Quantum Systems, Freie Universität, 14195 Berlin, Germany}
	\author{Clara Wassner}

	\affiliation{Dahlem Center for Complex Quantum Systems, Freie Universität, 14195 Berlin, Germany}
	\author{Giacomo Guarnieri}
        \affiliation{Department of Physics, University of Pavia, Via Bassi 6, 27100, Pavia, Italy}
	\affiliation{Dahlem Center for Complex Quantum Systems, Freie Universität, 14195 Berlin, Germany}
 
	\author{Jens Eisert}
	\affiliation{Dahlem Center for Complex Quantum Systems, Freie Universität, 14195 Berlin, Germany}
	\affiliation{Helmholtz Center Berlin, 14109 Berlin, Germany}
	
	\date{\today}

\begin{abstract}
Proving thermalization from the unitary evolution of a closed quantum system is one of the oldest questions that is still nowadays only partially resolved. Several efforts have led to various formulations of what is called the eigenstate thermalization hypothesis, which leads to thermalization under certain conditions on the initial states. These conditions, however, are highly sensitive on the precise formulation of the hypothesis. In this work, we focus on the important case of low entanglement initial states, which are operationally accessible in many natural physical settings, including experimental schemes for testing thermalization and for quantum simulation. We prove thermalization of these states under precise conditions that have operational significance. More specifically, motivated by arguments of unavoidable finite resolution, we define a random energy smoothing on local Hamiltonians that leads to local thermalization when the initial state has low entanglement. Finally we show that such a transformation affects neither the Gibbs state locally nor, under generic smoothness conditions on the spectrum, the short-time dynamics.

\end{abstract}
	
	\maketitle
	
Not long after their formulation, it became clear that the postulates of quantum mechanics were to a degree at odds with the principles of statistical mechanics \cite{vonNeumann29}. 
On the one hand, closed quantum systems evolve unitarily, thus preserving information about their initial conditions. On the other hand, statistical mechanics is centered around concepts such as irreversibility and thermalization, which allow to describe many-body systems in terms of just a few macroscopic quantities and parameters that depend neither on the initial conditions nor on the microscopic details ~\cite{jaynes1957informationI,jaynes1957informationII,1408.5148,ngupta_Silva_Vengalattore_2011,christian_review}. 
Both theories have nowadays provided correct predictions of uncountable experimental observations. Thermalization of closed many-body quantum systems is in particular observed to occur in practice with overwhelming evidence, also in quantum simulators which allow to probe their dynamics with high levels of precision \cite{Trotzky,Neill_2016, Kaufman_2016, Clos_2016,Somhorst2023,ShortGaussification}.
This apparent contradiction is eased once one only requires the expectation value of a subset of ``physical" observables, for example local observables, to agree with the predictions of statistical mechanics. As a matter of fact, in a lattice system, the state of the system reduced to local patches can relax to a thermal state while the global state evolves unitarily and remains pure. Nevertheless, a derivation of thermalization in this sense from the microscopic description of the dynamics has remained largely elusive. Efforts in this directions have led to various formulations of the \emph{eigenstate thermalization hypothesis} (ETH) \cite{Srednicki_1994,Deutsch_2018}, which posits that each eigenstate of a thermalizing Hamiltonian provides the same expectation values of physical observables as the ones provided by local thermal states. The ETH, however, is difficult to verify starting from a microscopic model. In addition, the precise conditions on the initial state and on the observables leading to thermalization are highly sensitive to the formulation of the hypothesis.
On the other hand, \emph{absence} of thermalization, as observed in integrable or localized systems \cite{Anderson_1958, Billy_2008, Abanin_2019}, seems to be the occurrence requiring particular, carefully set-up conditions. This has led to the idea of \emph{typicality}, i.e., that ``most" physical systems obey statistical mechanics. This usually means that if the system (for instance, its Hamiltonian) is drawn at random from a reasonable set, with overwhelming probability thermalization occurs as expected, and while any given system under examination is usually not randomly drawn, it should behave like a typical one unless explicitly engineered not to. Works in this direction \cite{Goldstein_2006,Popescu_2006,Cramer_2012, Reimann_2016, Dabelow_2022} have contributed to formalize the idea that thermalization is the rule, rather than the exception, even in isolated systems. 


A make or break question is left open, however: Why would one generically expect -- starting with natural initial states such as low-entanglement states -- quantum systems to relax to states that are operationally indistinguishable from Gibbs states by local measurements. This is the actual empirical observation of apparent thermalization that needs to be answered by a theoretical framework. The core prejudice to be overcome is not to prove thermalization for all Hamiltonians, but only for most such Hamiltonians that are slightly randomized, and still yield extremely similar dynamical predictions compared to the original Hamiltonian. 

In this work, we provide an important missing piece of the puzzle of proving thermalization of low-entanglement states. We show that initial low-entanglement states thermalize under appropriately weakly randomized local Hamiltonians. In contrast to previous approaches \cite{Cramer_2012, Reimann_2016}, we focus on making 
the randomization as weak as possible while still ensuring thermalization under the randomized Hamiltonian. Under a meaningful spectral smoothness assumption, the randomization can be chosen to be weak enough such that it does not affect the short-time dynamics. In addition, we show that the randomization does not affect the thermal properties of the Hamiltonian. With these last two points, we take steps towards bridging the gap between the spatially local description of the system given by a local Hamiltonian, and the energy-local description given by random matrix theory necessary to explain thermalization. Accepting that the Hamiltonian will anyway only be known to a finite precision, such a perturbative typicality argument goes a long way towards explaining why a dynamical convergence to Gibbs states is so ubiquitous in 
nature.

It is known that quantum lattice models that are prepared in typical states that are restricted to a narrow window centered around a given energy -- invoking notions of equivalence of ensembles -- will give rise to local expectation values that are indistinguishable from thermal ones \cite{brandao_2015}. While this is conceptually extremely insightful, it is not clear how such states can arise from physical dynamics of natural initial states.
What is, in contrast, very natural is to 
consider low-entanglement initial conditions and natural time evolution. Such low-entanglement states, in addition to being fundamental to the study of lattice systems, are commonly
the only accessible initial states to thermalization experiments in quantum simulators or numerical 
examination of thermalization, for example, in the context of probing many-body localization \cite{Bardarson2012, kim2014,Singh2016, BertoniMBL,Smith2016}, and are often 
considered in works on relaxation and equilibration \cite{Cramer_2008,Linden_etal09,Gluza_2019}.

\begin{figure}[h!]
    \centering
    \includegraphics[width=.9\linewidth]{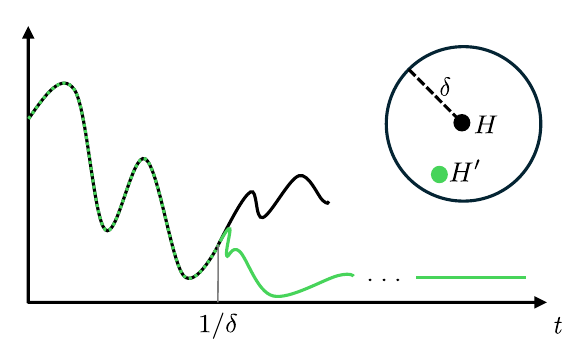}
    \caption{Illustration of the dynamics of the original and perturbed Hamiltonians. The distance of the new Hamiltonian from the original one is bounded by a parameter $\delta$. The plot shows how the evolution of an observable under $H$ or $H'$ behaves: the two evolutions agree up to a time $\sim 1/\delta$, and the perturbed Hamiltonian equilibrates to a thermal value.}
    \label{fig:summary}
\end{figure}

Concretely, given a local Hamiltonian $H$ and an initial low-entanglement state $\rho$, we define an ensemble of random Hamiltonians such that, if $H'$ is a Hamiltonian drawn from this ensemble: 1. the Gibbs states of $H$ and $H'$ are locally indistinguishable, and 2. with overwhelming probability the equilibrium state resulting from the unitary evolution $\ee^{-\ii H't}\rho \ee^{\ii H't}$ is locally indistinguishable from said Gibbs state. 
Importantly, we further give an assumption under which the dynamics of $H$ and $H'$ are indistinguishable in the short-time regime. (See \cref{fig:summary}).
We address this by studying states in an analog of the micro-canonical ensemble which is allowed to have support on the whole energy spectrum. Building upon and extending the work of 
Ref.~\cite{brandao_2015}, we show that, under precise and physically meaningful conditions, these states are locally equivalent to Gibbs states. We later show 
that such states arise from the \mbox{long-time} dynamics of low-entanglement initial states under typical Hamiltonians. Our typicality transformation, defining the ensemble of random Hamiltonians, consists of a unitary operation that randomly 
mixes eigenstates with nearby energies.
 Related transformations have been introduced in Ref.~\cite{Foini_2019} to compute higher order correlation functions under a generalized version of the ETH stemming from local (in energy) unitary invariance, and this has been connected to the theory of free probability in Ref.~\cite{Pappalardi_2022}. In our work, we explicitly demonstrate how such a transformation leads to equilibrium states being locally indistinguishable from thermal states. Moreover, by elucidating how the original Hamiltonian is related to the transformed one, we motivate a meaningful, operational interpretation of our Hamiltonian-typicality approach: 
Given a physical system in the right conditions, one can find a local Hamiltonian $H$ which well describes the dynamics. Nevertheless, there is a host of other models $H'$ which, if $H$ is well behaved, describe very similar dynamics as well as equilibrium properties, and are close to $H$. But $H$ cannot be known to infinite precision. This operational uncertainty can be modeled by our Hamiltonian ensemble.

\paragraph*{Physical setting, equilibration, and thermalization.}
Throughout this work we consider a cubic $D$-dimensional lattice $\Lambda$ with $N$ sites.  With each site $i$ we associate a Hilbert space $\mathcal H_i$ of dimension $d$. $H$ is a $k$-local Hamiltonian on $\Lambda $ for a constant $k$, meaning that it is of the form $H=\sum_{i\in \Lambda} h_i$ where $h_i$ has operator norm $1$ and is only supported on sites $j$ s.t.\ $\mathrm d(i,j)\leq k$. Here, $\mathrm d(\cdot,\cdot)$ is the standard Manhattan distance on the lattice. We will denote the spectral decomposition of $H$ as $H=\sum_\nu E_{\nu} P_\nu$. For ease of presentation, in the main text we will assume the eigenstates to be non-degenerate, i.e., $\tr{P_\nu}=1$ for all $E_\nu$. In the Supplemental Material, we prove our results relaxing this assumptions. We will use standard $O$ and $\Omega$ notation, and we will use $\tilde O, \tilde \Omega$  to denote asymptotic upper bounds where logarithmic factors are ignored, i.e. $f(h)=\tilde O(g(h))$ iff $f(h)=O(\log^r(h)g(h))$ for some $r$.
Before discussing thermalization, we need to discuss a natural prerequisite: equilibration \cite{christian_review,1110.5759,ReimannKastner12,PhysRevLett.123.200604}. Intuitively, a state equilibrates if after a finite amount of time the system reaches an invariant state. This means that if one looks at the 
whole evolution as time tends to infinity, the system will spend most of its time close to equilibrium, and hence the state at equilibrium should agree with the infinite-time averaged state
\begin{equation}\label{eq:infinite_state}
      \rho_\infty^H=\lim_{T\to\infty}\frac 1 T\int_0^T\mathrm dt \,\rho^H(t),
  \end{equation}
with $\rho^H(t)=\ee^{iHt}\rho \ee^{-iHt}$. The system is then said to equilibrate if the fluctuations around the average for an observable $A$ are small in the sense of
\begin{equation}
    \Delta A_\infty:=\lim_{T\to\infty}\frac 1 T\int_0^T\mathrm dt\, \tr{A(\rho^H(t)-\rho_\infty^H)}^2\xrightarrow{N\to\infty} 0.
\end{equation}
Equilibration in this sense has been rigorously proven in a variety of settings \cite{Farrelly_2017, PhysRevLett.123.200604, Haferkamp_2021}.
Thermalization essentially consists of a stronger requirement of equilibration, where the equilibrium state must coincide with the thermal (Gibbs) state
\begin{equation}
g_\beta(H)=\frac{\ee^{-\beta H}}{Z},\quad Z=\tr{\ee^{-\beta H}}.
\end{equation}
In particular, we will be interested in 
understanding how distinguishable a given state is from a Gibbs state, given only access to local observables. We introduce the quantity
 \begin{equation}
           D_l(\rho, \sigma):=\frac{1}{|\mathcal C_l|}\sum_{C\in\mathcal C_l}\|\rho_C-\sigma_C\|_1,
\end{equation}
where $\mathcal C_l$ denotes the set of all hypercubes in the lattice of side length $l$ and $\rho_C$ is the state $\rho$ reduced to $C$. 
If $\rho,\sigma$ are translationally invariant, this is simply equal to the trace norm $\|\rho_C-\sigma_C\|_1$ for any $C\in\mathcal C_l$. Otherwise it measures the distinguishability of $\rho$ and $\sigma$ given access to observables of the form $\frac{1}{|\mathcal C_l|}\sum_{C\in\mathcal C_l} O_C$, such as typical average local quantities in statistical physics.
We will then say that a state is \emph{locally thermal} if $D_l(\rho,g_\beta(H))$ converges to $0$ with $N$ for some $\beta$.
In the remainder of this work, we will make some assumptions on the Gibbs state of the system under examination, $g_\beta(H)$. First, we will require it to have exponential decay of correlations. A state $\rho$ is said to have \emph{exponential decay of correlations} if for some $\xi>0$, for any two regions $X,Y\subset \Lambda$ and $A,B$ supported on $X$ and $Y$ respectively,
		\begin{equation}
		 |\langle AB\rangle_\rho-\langle A\rangle_\rho\langle B\rangle_\rho|\leq \|A\|\|B\|\ee^{-\mathrm d(X,Y)/\xi} .
		\end{equation}
The Gibbs state $g_\beta(H)$ always has an exponential decay of correlations in one dimension \cite{Araki_1969}, as well as in any higher dimension above a certain critical temperature \cite{Kliesch_2014} that depends only on a few parameters of the system.
We will denote by $\sigma$ the standard deviation of the energy, i.e., $\sigma^2=\tr{g_\beta(H)H^2}-\tr{g_\beta(H)H}^2$, and assume $\sigma^2\geq \Omega(N)$, which implies that the specific heat capacity is non-zero in the thermodynamic limit.
    
 \paragraph*{Conditions for thermality.}
    The \emph{micro-canonical ensemble} is commonly defined as the maximally mixed state supported in a narrow window around a fixed energy. This physically models maximal uncertainty with an energy constraint. However, arbitrary and even physically motivated states such as low-entanglement states, can have non-trivial support over the whole spectrum during the whole time evolution; these states will therefore never be micro-canonical in the sense above, particularly not at equilibrium. With the goal of overcoming this issue, we leverage techniques used to prove equivalence between micro-canonical and canonical ensemble. 
    It has been proven in Ref.\ \cite{brandao_2015} that equivalence with a canonical ensemble is already achieved by states confined in a micro-canonical window that are not maximally mixed, but have sufficiently high entropy. We adapt this result to the situation in which the state is not confined in a window, but instead has a support over many such windows plus decaying tails. First of all, we call this a \emph{generalized micro-canonical ensemble}, meant to capture the thermal behavior of states that are supported on regions of the spectrum larger than what the usual micro-canonical ensemble allows.
    
\begin{figure}[ht]
        \centering
        \includegraphics[width=.5\textwidth]{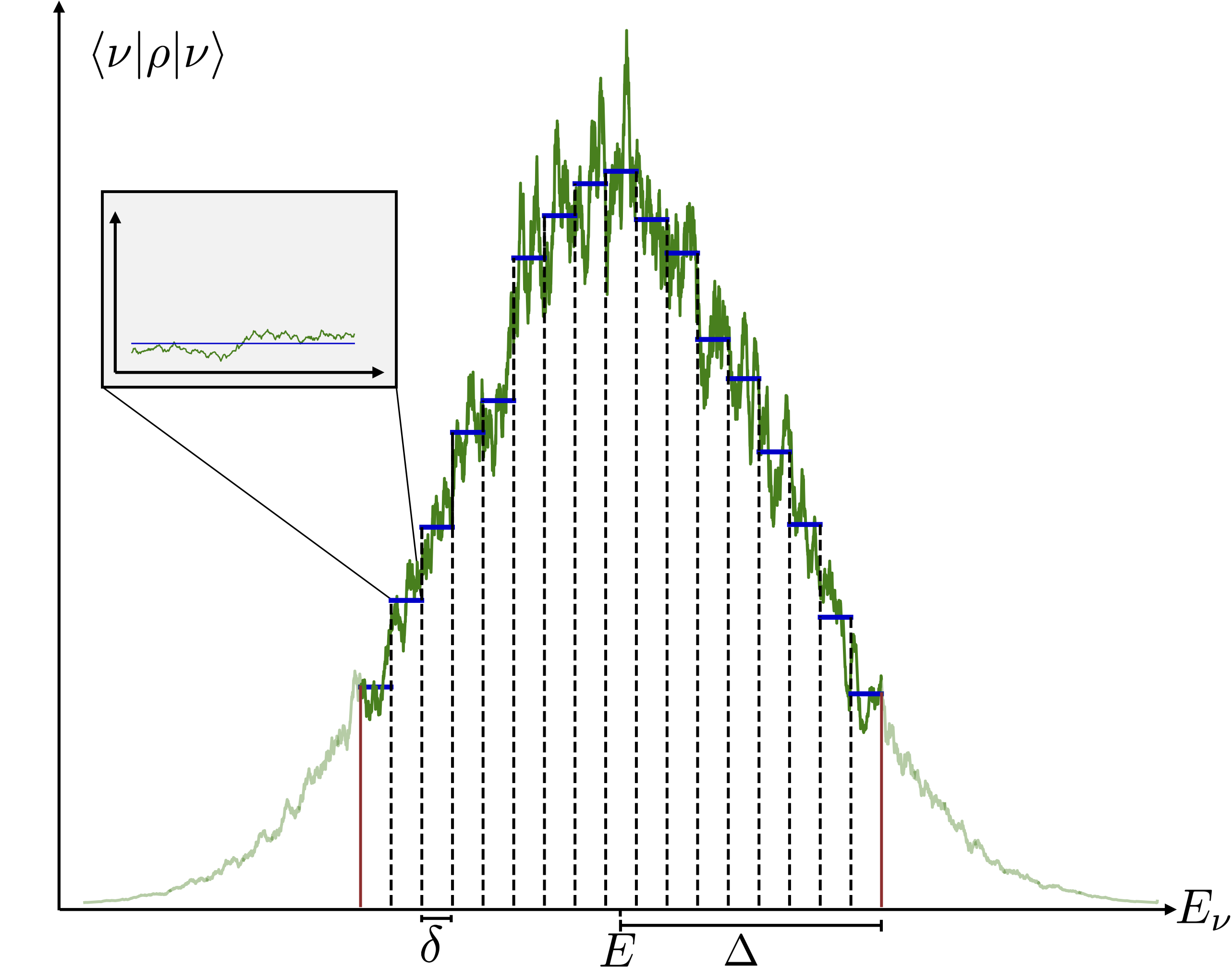}
        \caption{Cartoon illustration of the setting of \cref{def:gme} and \cref{def:agme}. In this example, the state is diagonal in the energy eigenbasis and it is represented as a probability distribution. The blue line is a GmE state, and the green line an approximate GmE state. The insert shows both states restricted to one of the windows.}
        \label{fig:figure}
    \end{figure}

\begin{definition}[Generalized micro-canonical ensemble (GmE)]\label{def:gme}
Let $[E-\Delta, E+\Delta]$ ($\Delta >0$) denote an energy window centered around a value $E$ and divided into $K$ bins of various size $\delta_k$, with $k=1,\dots, K$.
Let $\bgreek{\delta} = (\delta_1,\ldots ,\delta_K)$; we define a \emph{generalized micro-canonical ensemble} (GmE) to be the state of the form
\begin{equation}
    \omega:=\omega(E,\Delta, \bgreek{\delta},\mathbf{q})=\sum_{k=1}^K q_k \omega_{\delta_k}
\end{equation}
where $\omega_{\delta_k}$ is the micro-canonical ensemble supported inside the window $k$, and where $\mathbf q=(q_1,\ldots, q_K)$ such that $\sum_{k=1}^K q_k =1$.
\end{definition}
This state therefore physically represents a statistical combination of micro-canonical ensembles; see \cref{fig:figure}.
For the sake of simplicity, we will choose $\delta_k = \delta$ from now on. 
However, in the Supplemental Material we show that all our results still hold true if this assumption is relaxed. Before stating our first main result, we need to introduce the notion of the \emph{Berry-Esseen} (BE) error, which quantifies the difference between a state written in the energy eigenbasis and a Gaussian distribution. More specifically, if $\Pi_x$ is the projector onto all energy eigenstates with energy smaller than $x$, then the BE error of $\rho$ with respect to $H$ is defined as $\zeta_N=\sup_x|\tr{\rho\Pi_x}-G(x)|$, where $G(x)$ is the Gaussian distribution with the same mean and variance as $\rho$.
It was proven that if $\rho$ has exponential decay of correlations, then $\zeta_N\leq \tilde O(N^{-1/2})$ \cite{brandao_2015}. Simple examples saturating this bound are known, and this bound is expected to be saturated by certain non-thermalizing models. Nonetheless, under some more generic constraints, such as highly entangled eigenstates, a more favorable scaling is expected, even up to $\zeta_N\leq \ee^{-\Omega(N)}$ \cite{Rai_2023}. From now on we denote by $\zeta_N$ the BE error with $\rho=g_\beta(H)$. In what follows, we will assume $\zeta_N\leq \tilde O(N^{-1/2-\kappa})$ for some $\kappa\geq 0$. This includes the worst case $\kappa=0$.

\begin{theorem}[Ensemble equivalence]\label{thm:ensemble_equivalence}
    Let $H$ be a local Hamiltonian and $\beta$ an inverse temperature for which the Gibbs state $g_\beta(H)$ has exponential decay of correlations, standard deviation $\sigma\geq \Omega(\sqrt{N})$ and Berry-Esseen error $\zeta_N\leq \tilde O(N^{-1/2-\kappa})$ for $\kappa\geq 0$. Let $\omega$ denote a GmE with $\Delta, \delta$ satisfying 
        \begin{equation}
        \begin{aligned}               &\ee^{\Delta^2/\sigma^2}\leq \tilde O\left(N^{\frac{1-\alpha}{D+1}}\right),\quad  \Omega\left(N^{\frac{1-\alpha}{D+1}-\kappa}\right)\leq \delta\leq \sigma,
        \end{aligned}
        \end{equation} 
        with $\alpha\in[0,1)$ and such that $|E-E_\beta|\leq \sigma$. Then for any side length $l$ such that $l^D \leq C_1\, N^{\frac{1}{D+1}-\gamma_1\alpha}$, the following holds
       \begin{equation}
            D_l(\omega, g_\beta(H))\leq C_2 N^{-\gamma_2\alpha},
        \end{equation}
        with $C_1, C_2$ being system-size independent constants, and $\gamma_1,\gamma_2$ only depend on the dimension of the lattice $D$.
\end{theorem}
This first main result shows that, for appropriate choices $\Delta$ and $\delta$, GmE states are locally indistinguishable from Gibbs states. A GmE state can be seen as a mixture of micro-canonical states spanning a range of temperatures and \cref{thm:ensemble_equivalence} shows that as long as its range is small enough, the state still looks thermal with a well-defined temperature. Notice that if $\kappa>0$, i.e., if the BE error is better than the worst case scenario, $\delta$ can be chosen to decay with the system size. 

Keeping in mind our initial goal of capturing equilibrium states resulting from natural and physically motivated initial states, it may seem artificial to consider only block-like states with sharp jumps between energy intervals. Therefore, postponing the discussion about their physicality to the next Section below, we first of all prove that the same ensemble equivalence holds if the state's structure gets more relaxed, i.e., if it is only approximately GmE in the sense precisely elucidated below.
\begin{definition}[Approximate GmE]\label{def:agme}
    Let $E, \Delta,\bgreek{\delta},\mathbf q$ be as in \cref{def:gme}. We define $\omega_\eta$ an approximate GmE if it is of the form
    \begin{equation}\label{eq:agme}
        \omega_\eta=p_\Delta\left(\sum_{k=1}^K q_k \tilde\omega_{\delta_k}\right) + (1-p_\Delta)\rho_{\mathrm{tail}} 
    \end{equation}
    and its von Neumann entropy satisfies
    \begin{equation}
        \sum_{k=1}^{K} q_k (S(\omega_{\delta_k})-S(\tilde\omega_{\delta_k}))\leq \eta,
    \end{equation}
    with $\tilde\omega_{\delta_k}$ being defined on the Hilbert space spanned by the eigenstates in the $k$-th energy bin, and $\rho_{\mathrm{tail}}$ on the Hilbert space spanned by the eigenstates outside $[E-\Delta, E+\Delta]$.  
\end{definition}
This state represents a more physical version of a GmE state inside the energy window $\Delta$, with decaying tails outside, that has an entropy $\eta$-close to the maximum one.
Importantly, in the Supplemental Material we demonstrate that \cref{thm:ensemble_equivalence} holds true also for the approximate GmE and takes the form
\begin{equation}\label{eq:main1}
    D_l(\omega_\eta, g_\beta(H))\leq C_2 N^{-\gamma_2\alpha} +  2(1-p_\Delta),
\end{equation}
with $\eta \leq N^{\frac{1-\alpha}{D+1}}$.

    This shows that states which are concentrated around an energy regime and are sufficiently ``smooth" are locally equivalent to Gibbs states.  
    The results above are a generalization of the equivalence of ensembles result of Ref.\ \cite{brandao_2015}, and their proof is presented in the Supplemental Material, Section I.
    
    \paragraph*{Thermalization via energy smoothing.}\label{sec:smoothing}
    The next question is whether (approximate) GmE states can  actually be obtained from Hamiltonian evolution of isolated systems under natural, or typical, conditions. There are two main aspects we consider when talking about ``natural" conditions: (i) the Hamiltonian responsible for the time evolution; (ii) the initial state of the system. Regarding (i), we consider \emph{typical} Hamiltonians in a sense that we will make precise below, in order to exclude edge cases or fine-tuned Hamiltonians for which one does not expect thermalization (for instance, integrable models). Concerning (ii), previous typicality approaches have assumed the initial state to be confined in a well-defined energy interval, and have shown properties of the relaxation towards a micro-canonical ensemble in said interval. Here, instead, we start from the assumption of exponential decay of correlations, i.e., low entanglement between spatially separated regions, which we take as natural starting states for lattice systems. These states have been shown to have fast decaying tails in energy \cite{Anshu_2016} which makes them ideal candidates to flow to approximately GmE states. 
    Here, for simplicity of presentation, we focus on the case of product states, and leave the more general case of states with exponential decay of correlations and the proofs to the Supplemental Material, Section II.

    Let us expand on the ensemble of typical Hamiltonians that we consider. Starting from any local Hamiltonian on the lattice, we divide its energy spectrum into energy intervals of equal width $\delta$ which we call $I_k$, for $k=1,\cdots, K$. The eigenstates contained within each interval span a vector space which we call $\mathcal W_k$. We then consider unitaries of the form $U=\bigoplus_k U_k$, where $U_k$ is drawn from the Haar measure of the unitary group acting on $\mathcal W_k$. This defines an ensemble of random unitaries which we denote as $\mathcal E(\delta)$. A typical Hamiltonian is then $UHU^\dagger$ for such a random unitary $U$. All these Hamiltonians have the same spectrum as the original local Hamiltonian $H$, and the randomization given by $U$ is designed to preserve the expected energy of any state. 

    The following is a consequence of measure concentration and the results of 
    Ref.\ \cite{Anshu_2016} about energy tails of product states.
    
    \begin{lemma}[Approximate GmE at equilibrium]\label{lem:high_entropy}
        Let $\rho$ be a product state and $H$ be a local Hamiltonian. Let $\rho_\infty^{UHU^\dagger}$ be defined as 
        in \cref{eq:infinite_state}, 
        where $U$ is drawn from $\mathcal E(\delta)$. Consider the interval $I=[E-\Delta,E+\Delta]$ around $E=\tr{\rho UHU^\dagger}$  with $\Delta\geq \omega(\sqrt{N})$ an integer multiple of $\delta$, then
        \begin{equation}
            \rho_\infty^{UHU^\dagger}=p_\Delta\left(\sum_{k:I_k \subset I} q_k \tilde\omega_{\delta_k}\right) + (1-p_\Delta)\rho_{\mathrm{tail}} 
        \end{equation}
        with $p_\Delta\geq 1-\ee^{-c_1\frac{\Delta^2}{N}}$, and for $r>0,$ with probability at least $1-2^{-r+1}$, we have
          \begin{equation}
        \sum_{k:I_k \subset I}q_k (S(\omega_{\delta_k})-S(\tilde\omega_{\delta_k}))\leq r,
    \end{equation}
    where $c_1$ is a system-size independent constant.
    \end{lemma}
    We have then the following consequence on typical 
    thermalization.
    
\begin{theorem}[Typical thermalization]
    \label{thm:thermalization}
        Let $H$ be a $k$-local Hamiltonian and $\rho$ be a product state. Let $g_\beta(H)$ be the Gibbs state of $H$ at inverse temperature $\beta$ such that $|\tr{g_\beta(H)H}-\tr{\rho H}|\leq \sigma $. Assume $g_\beta(H)$ has exponential decay of correlations, $\sigma \geq \Omega(\sqrt N)$, and $\zeta_N\leq \tilde O(N^{-1/2-\kappa})$.
        For any constant $\alpha\in [0,1)$, choosing $\delta=\Omega\left(N^{\frac{1-\alpha}{D+1}-\kappa}\right)$, with probability at least $1-\exp( -c_2 N^{\frac{1-\alpha}{D+1}})$ drawing $U$ at random from $\mathcal E(\delta)$, we have 
        \begin{equation}
 D_l(\rho_\infty^{UHU^\dagger},g_\beta(H))\leq C_2\,N^{-\gamma_2\alpha}+ \tilde O\left(N^{-\gamma_3(1-\alpha)}\right),
        \end{equation}
        where $c_2, C_2,\gamma_2,\gamma_3$ are system-size independent constants.
    \end{theorem}
    In the Supplemental Material, Section II, we state and prove these results more generally for any state concentrated around its average, which includes states with exponentially decaying correlations. The consequences of this relaxation of the assumption is that the decay in the system size is slower. \cref{thm:thermalization} shows that the equilibrium state is locally thermal; in the Supplemental Material, Section II, we show under mild spectral assumptions that the randomized Hamiltonian equilibrates with high probability to this state. 
   Although it may seem strange at first glance that under the dynamics of $UHU^\dagger$ the state thermalizes to the Gibbs state of $H$ and of $UHU^\dagger$, 
   we prove that the Gibbs states of these two Hamiltonians are locally indistinguishable. More specifically,  under the same assumptions as \cref{thm:thermalization}, for any $U$ drawn from $\mathcal E(\delta)$ we have
      \begin{equation}D_l(g_\beta(H),g_\beta(UHU^\dagger))\leq  O(N^{-\gamma_4\alpha-\gamma_4\kappa})
        \end{equation}
        for system-size independent constants $\gamma_4,\gamma_5$.
The proof may be found in the Supplemental Material, Section II, and easily generalizes to other choices of $\delta$. As anticipated, the unitary ensemble $\mathcal E(\delta)$ is chosen in order to approximately preserve the energy of any state; this implies that $U\sim \mathcal E(\delta)$ approximately commutes with the Hamiltonian, and we show $\|H-UHU^\dagger\|_\infty\leq \delta$. For the choice of $\delta$ as in \cref{thm:thermalization} we derive the following result
\begin{equation}\label{eq:time_indist}
    \|\ee^{-iHt}\rho \ee^{iHt}-\ee^{-iH' t}\rho \ee^{iH' t}\|_1\leq 2t\, O\left(N^{\frac{1-\alpha}{D+1}-\kappa}\right),
\end{equation}
with $H'=UHU^\dagger$. This means that the dynamics under $H$ and $UHU^\dagger$ are indistinguishable up to a time $t^*\sim N^{\kappa-\frac{1-\alpha}{D+1}}$. If $\kappa>0$, that is the BE error decays faster than the worst-case scenario, $\alpha$ can be chosen such that $t^*$ increases with the system size. In other words $H$ and $UHU^\dagger$ generate nearly the same dynamics for a time $t^*\sim \mathrm{poly}(N)$. Finally, we would like to emphasize that our rigorous approach allows to put on precise and solid ground some of the results obtained on equilibration in Ref.~\cite{reimann_transportless_2019}.

It is now worth noting that if both $\rho$ and $\sigma$ are translation-invariant, the averaging over regions in the definition of $D_l(\rho,\sigma)$ can be dropped, making the indistinguishability statement valid for any observable supported on an individual small region. We show that if the original Hamiltonian is translation-invariant, we recover this property to some extent in the equilibrium state of the perturbed Hamiltonian. More specifically, consider an observable $A$ supported in $C\in \mathcal C_l$ and $H$ translation-invariant. For $U$ drawn from $\mathcal E(\delta)$ we show, {assuming all windows centered around an extensive energy contain exponentially many eigenvalues, that except with probability $\ee^{-\Omega(N)} $ ,
    \begin{equation}\label{eq:translinv}\begin{aligned}
        \bigg|\mathrm{tr}\bigg(\big(g_\beta(H)-\rho_\infty&^{UHU^\dagger }\big) A\bigg)\bigg|\leq \\& \ee^{-\Omega(N)}+D_l\left(\rho_\infty^{UHU^\dagger}, g_\beta(H)\right).
        \end{aligned}
    \end{equation}}
Details and proofs are available in the Supplemental Material, Section V.  By applying this to $\rho=g_\beta(UHU^\dagger)$ we also get translation invariance in the same sense for the randomized Gibbs state, that is, $\rho_\infty^{UHU^\dagger}$ can be replaced by $g_\beta(UHU^\dagger)$ in \cref{eq:translinv}.

\paragraph*{Dynamical thermalization.}
Turning to notions of \emph{dynamical thermalization},
we have investigated the typical time-evolution of the expectation value of a generic observable $A$, $\langle A \rangle_\rho := \tr{A\rho}$, under the evolution generated by $UHU^\dagger$. 
In the Supplemental Material, Section IV, {we show, assuming all windows centered around an extensive energy contain exponentially many eigenvalues, that except with 
probability $(N/\delta)^2\ee^{-\Omega(N)}$, the time-evolution is bounded by 
\begin{equation}
    |\langle A \rangle_{\rho^{UHU^\dagger}(t)}-\langle A \rangle_{\rho_\infty^{UHU^\dagger}}|\leq \ee^{-\Omega(N)} + R(t)  ,
\end{equation}}
where $R(t)$ is a function of $t$ depending on details of the spectrum of $H$, on $A$, and on $\rho$. Performing a similar analysis to the one of Ref.\ \cite{Reimann_2016} to our ensemble, and assuming that the spectrum in each window can be well approximated by a 
suitably flat
continuous spectrum, 
we show that 
\begin{equation}
    R(t)\sim \|A\|_\infty\frac{N^2}{\delta^2 t^2}.
\end{equation}
Hence, under this physical assumption, thermalization up to some $\epsilon$ is reached after a time $\sim N^{O(1)}/\epsilon$.

 \paragraph*{Discussion and conclusion.}
In this work, we have made substantial progress in the long-standing research quest of proving thermalization from first principles of quantum mechanics, by showing  thermalization of low-entanglement states under typical Hamiltonians. 
In particular, we have defined ensembles which naturally emerge from the time-evolution of low-entanglement states under typical Hamiltonian evolution and show that they are locally indistinguishable from the Gibbs ensemble. The typicality in the Hamiltonian is given by randomizing the eigenbasis locally in energy, and we show that this randomization does not affect the Gibbs state locally. Furthermore, we show that if the Gibbs state has a relatively generic smoothness property, the randomization does not affect the short-time dynamics in any discernible way. 

On a higher level, this is one of the main physical messages of this work: it is unrealistic to assume that a Hamiltonian can be specified up to arbitrary precision. The ubiquity of thermalization may then be explained by considering that its absence requires instead a high precision in the specification of the system. In this light, proving thermalization for all, or a large class of, local Hamiltonians might not be required for explaining this observation. 

The technical results established here are also expected to be helpful in the design of \emph{quantum algorithms} for preparing Gibbs states \cite{RouzeGibbs,BrandaoKastoryanoGibbs,Riera-PRL-2012}.
Relaxing the assumption of requiring the entire state to be globally indistinguishable from a Gibbs state (as suggested in Ref.\ \cite{Riera-PRL-2012}) may well be helpful in this endeavour.  

Several natural open questions remain. Importantly, the problem of identifying precise properties of the Hamiltonian achieving a lower than worst case BE error. For these Hamiltonians, our result predicts that a vast ensemble of other Hamiltonians exists that all have indistinguishable short-time dynamics and which eventually thermalize. Furthermore, while our randomization achieves thermalization, 
more work may be needed about exploring the precise
physical mechanisms that can be held responsible for them.
On the one hand, being able to implement such transformations in a controlled way might have implications in the form of algorithms for Gibbs state preparation. On the other hand, understanding under what condition an uncertainty in either state or Hamiltonian preparation can translate into a random energy-preserving perturbation such as the one we defined might shed light on the stability of many-body localization under realistic conditions. It is the hope that this work contributes to understanding that under a ``trembling hand'', most systems follow the laws of quantum statistical mechanics.


\paragraph*{Acknowledgements.} We thank Henrik Wilming and Ingo Roth on discussions during early stages of this project. 
This work has been supported by the DFG (FOR 2724 and CRC 183), the BMBF (MUNIQCAtoms), the Quantum Flagship (PasQuans2), the FQXi, and the ERC (DebuQC).

	\bibliography{bibRef.bib}
 
\newpage

\pagebreak
\clearpage
\foreach \x in {1,...,\the\pdflastximagepages}
{
	\clearpage
	

\section*{Supplementary material}

# Supplemental Material

## I. EQUIVALENCE OF ENSEMBLES

The main purpose of this section is to prove Theorem 1 of the main text in a more general version. For what follows, we will need two auxiliary lemmas, both stated and proven in ref. [1], the final result will follow from a similar argument to the one used to prove equivalence of statistical mechanics ensembles in the same work. The first one states that a state close in relative entropy to some other state with exponential decay of correlations will also be locally indistinguishable from that state.

**Lemma 1** ([1], Proposition 2). *Let $\rho$ be a state with correlation length $\xi$ and let $\tau$ be a state. Let $\mathcal{C}_l$ be the set of hypercubes in $\Lambda$ with edge length $l$, if for $\epsilon > 0$*

$$S(\tau||\rho) + 3 + \epsilon^{\frac{D+1}{D+2}} \frac{2\xi \ln(d)l^D + l + 2}{\xi \ln(2)} + \epsilon^{\frac{D+1}{D+2}} \log(N) \leq \epsilon \left(\frac{N}{\ln(4)^D \xi^D}\right)^{\frac{1}{D+1}} \tag{1}$$

*then*

$$D_l(\tau, \rho) \leq 7\sqrt{\epsilon^{\frac{D+1}{D+2}}}. \tag{2}$$

This theorem has important consequences. In less formal terms, it implies that bounding the relative entropy between the two states bounds their local distinguishability, assuming some bounds on $\epsilon$ and the local regions:

**Corollary 1** (Local indistinguishability from relative entropy closeness). *Let $\rho$ be a state with correlation length $\xi$ and let $\tau$ be a state. Suppose that for some $\epsilon$, we have*

$$S(\tau||\rho) \leq \epsilon N^{\frac{1}{D+1}}. \tag{3}$$

*Let*

$$\tilde{\epsilon} := \frac{4}{(\xi \ln(4))^{\frac{D}{D+1}}} \epsilon. \tag{4}$$

*If*

$$\tilde{\epsilon} \geq \left(\frac{\ln(4)^D \xi^D}{N}\right)^{\frac{D+2}{D+1}} (4 \log(N))^{D+2} \tag{5}$$

*then for*

$$l^D \leq \frac{1}{4} \frac{\xi \ln(2)}{2\xi \ln(d) + 3} \tilde{\epsilon}^{\frac{1}{D+2}} \left(\frac{N}{\xi^D \ln(4)^D}\right)^{\frac{1}{D+1}} \tag{6}$$

*we have*

$$D_l(\tau, \rho) \leq 7\sqrt{\tilde{\epsilon}^{\frac{D+1}{D+2}}}. \tag{7}$$

The second lemma is a variant of the Berry-Esseen theorem for quantum lattice systems. We denote with $\{|\nu\rangle\}_\nu$ and $\{E_\nu\}_\nu$ the eigenstates and corresponding eigenvalues of a local Hamiltonian $H$. If a state $\rho$ is uncorrelated, the energy probability distribution $\langle \nu|\rho|\nu\rangle$ behaves like a sum of independent random variables, and converges to a normal distribution with the system size. The Berry-Esseen theorem for quantum lattice systems gives system-size dependent bounds on the speed of convergence of the distribution.

**Definition 1** (Berry-Esseen error). *Let $\rho$ be a state and $H$ a local Hamiltonian, define $\mu := \operatorname{tr}(\rho H)$ and $\sigma^2 = \operatorname{tr}(\rho(H - \mu)^2)$ and let the functions $F$ and $G$ be defined as*

$$F(x) = \sum_{\nu: E_\nu \leq x} \langle \nu|\rho|\nu\rangle, \qquad G(x) = \frac{1}{\sqrt{2\pi\sigma^2}} \int_{-\infty}^{x} e^{-\frac{(y-x)^2}{2\sigma^2}}, \tag{8}$$

*then the Berry-Esseen error is the supremum of the difference of these two cumulative distribution functions*

$$\zeta_N(\rho, H) = \sup_{x \in \mathbb{R}} |F(x) - G(x)|. \tag{9}$$

**Lemma 2** ([1], Lemma 8). *Suppose $g_\beta(H) = \frac{e^{-\beta H}}{Z(\beta)}$ has exponential decay of correlations, let $\sigma^2 = \operatorname{tr}(g_\beta(H)H^2) - \operatorname{tr}(g_\beta(H)H)^2$ and assume $\sigma^2 \geq \Omega(N)$. Then*

$$\zeta_N(g_\beta(H), H) \leq R \frac{\ln(N)^{2D}}{\sigma} \tag{10}$$

*where $R$ is a constant depending on the correlation length, the locality of the Hamiltonian, and the dimension of the lattice.*

A tighter bound, without the logarithmic factor, holds for Gibbs states at sufficiently high temperature [2]. Throughout the document, we define $\zeta_N := \zeta_N(g_\beta(H), H)$. We are now ready to prove the following statement. This is a more general version of the main theorem of [1] for the case of the GmE and the proof follows from appropriately adapting the argument in that work.

**Theorem 1** (Thermality condition). *Let $H$ be a local Hamiltonian and let*

$$\frac{1}{6\sqrt{2\pi}} \sigma N^{\frac{-1}{D+1}} \geq \epsilon \geq \max\left\{ \log\left(\frac{1}{\tilde{\zeta}_N}\right), e^{1+\sqrt{5}} \ln(N)^{2D} \right\} N^{\frac{-1}{D+1}},$$
$$3\sqrt{2\pi}\, e^2 \sigma \tilde{\zeta}_N \leq \Delta \leq \sigma \sqrt{\ln\left(\frac{\epsilon N^{\frac{1}{D+1}}}{R \ln(N)^{2D}}\right)}, \tag{11}$$

*for some $\tilde{\zeta}_N \geq \zeta_N$ such that $\sigma\tilde{\zeta}_N \leq R\ln(N)^{2D}$. Let $\tau$ be a state supported only on eigenstates with energy in $I = [E+\Delta, E-\Delta]$ for some energy $E$. Divide the interval $I$ into subintervals $[e_k, e_{k+1})$ and suppose $\delta_k = |e_{k+1} - e_k|$ satisfies*

$$\delta_k^* := 3\sqrt{2\pi}\, e^{\frac{1}{2}\left(\frac{\Delta(k)}{\sigma}+2\right)^2} \sigma \tilde{\zeta}_N \leq \delta_k \leq \sigma \tag{12}$$

*where we have defined*

$$\Delta(k) := \begin{cases} |e_k - E| & \text{if } E < e_k, \\ |e_{k+1} - E| & \text{if } E > e_{k+1}, \\ 0 & \text{if } E \in [e_k, e_{k+1}]. \end{cases} \tag{13}$$

*Call the projector onto the eigenstates contained in the k-th window $\Pi_k$. If for each $k$*

$$S(\tau) \geq \sum_k \operatorname{tr}(\Pi_k \tau) \log(\operatorname{tr}(\Pi_k)) - \epsilon N^{\frac{1}{D+1}}, \tag{14}$$

*then for any $l$ such that*

$$l^D \leq \frac{1}{4}\left(\frac{4}{(\xi \ln(4))^{\frac{D}{D+1}}}\right)^{\frac{1}{D+2}} \frac{\xi \ln(2)}{2\xi \ln(d) + 3} \epsilon^{\frac{1}{D+2}} \left(\frac{N}{\xi^D \ln(4)^D}\right)^{\frac{1}{D+1}}, \tag{15}$$

*we have*

$$D_l(\rho, g_\beta(H)) \leq 7 \left(\frac{4(\beta \log(e) 3\sqrt{2\pi} + 1)}{(\xi \ln(4))^{\frac{D}{D+1}}}\right)^{\frac{1}{2}\frac{D+1}{D+2}} \epsilon^{\frac{1}{2}\frac{D+1}{D+2}} \tag{16}$$

*for any $\beta$ such that $|E - \operatorname{tr}(H g_\beta(H))| \leq \sigma$ and $g_\beta(H)$ has exponential decay of correlations.*

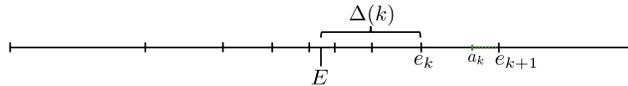

FIG. 1. A sketch of the interval $\Delta$ and the sub-intervals $\delta_k$ appearing in Theorem 1.

*Proof.* We would like to apply Lemma 1. We then need to bound the relative entropy $S(\tau||g_\beta(H))$. We begin by writing, by definition of relative entropy (here log denotes the logarithm in base 2),

$$S(\tau||g_\beta(H)) = -S(\tau) + \beta \log(e) \text{tr}(H\tau) + \log(Z(\beta)). \tag{17}$$

In a next step, we have

$$\text{tr}(H\tau) = \sum_k \text{tr}(\Pi_k H \Pi_k \tau) \leq \sum_k \text{tr}(\tau_k) e_{k+1}, \tag{18}$$

where we have defined $\tau_k := \Pi_k \tau \Pi_k$ and have used that $\Pi_k H \Pi_k \leq e_{k+1}$. We can then conclude that

$$S(\tau||g_\beta(H)) \leq -S(\tau) + \sum_k \text{tr}(\tau_k)(\log(e)\beta e_{k+1} + \log(Z(\beta))). \tag{19}$$

We define

$$Z_k(\beta) := \sum_{E_\nu \in [a_k, e_{k+1}]} e^{-\beta E_\nu} \tag{20}$$

for some $a_k \geq e_k$ to be defined later. Then we write for each $k$,

$$Z(\beta) = \frac{Z(\beta)}{Z_k(\beta)} Z_k(\beta). \tag{21}$$

At this point, notice that

$$Z_k(\beta) = \sum_{E_\nu \in [a_k, e_{k+1}]} e^{-\beta E_\nu} \leq d_k e^{-\beta a_k}, \tag{22}$$

where we have defined $d_k := \text{tr}(\Pi_k)$. We then have

$$S(\tau||g_\beta(H)) \leq -S(\tau) + \sum_k \text{tr}(\tau_k) \log(d_k) + \sum_k \text{tr}(\tau_k) \log\left(\frac{Z(\beta)}{Z_k(\beta)}\right) + \beta \log(e) \sum_k \text{tr}(\tau_k) \tilde{\delta}_k, \tag{23}$$

with $\tilde{\delta}_k := e_{k+1} - a_k$. We now need to show that $Z(\beta)/Z_k(\beta)$ is not too large for an appropriate choice of intervals. To this end, we write

$$\frac{Z_k(\beta)}{Z(\beta)} = \sum_{E_\nu \in [a_k, e_{k+1}]} \langle \nu|g_\beta(H)|\nu\rangle. \tag{24}$$

We proceed by using that the energy distribution of a thermal state is close to a Gaussian (Lemma 2). Define the functions $F$ and $G$ as

$$F(x) := \sum_{E_\nu \leq x} \langle \nu|g_\beta(H)|\nu\rangle, \qquad G(x) = \frac{1}{\sqrt{2\pi\sigma^2}} \int_{-\infty}^x dx\, e^{-\frac{(x-E_\beta)^2}{2\sigma^2}} \tag{25}$$

with $E_\beta := \text{tr}(H g_\beta(H))$. Then

$$\begin{aligned}\frac{Z_k(\beta)}{Z(\beta)} &= F(e_{k+1}) - F(a_k) \geq G(e_{k+1}) - G(a_k) - 2\sup|F(x) - G(x)| \\ &\geq G(e_{k+1}) - G(a_k) - 2\zeta_N.\end{aligned} \tag{26}$$

Recall that we wish to lower bound $Z_k(\beta)/Z(\beta)$. If $\tilde{\delta}_k$ is too small, $G(e_{k+1}) - G(a_k)$ will also be small and the lower bound in the above equation will be negative, hence we need to pick $\tilde{\delta}_k$ large enough such that $G(e_{k+1}) - G(a_k) \geq 3\tilde{\zeta}_N$ for some $\tilde{\zeta}_N \geq \zeta_N$ such that $\sigma\tilde{\zeta}_N \leq R\ln(N)^{2D}$, which will yield $Z_k(\beta)/Z(\beta) \geq \tilde{\zeta}_N$. The reason why we introduce $\tilde{\zeta}_N$ is that if $\zeta_N$ itself is too small, the lower bound $Z_k(\beta)/Z(\beta) \geq \zeta_N$ might be too loose later. We have

$$G(e_{k+1}) - G(a_k) = \frac{1}{\sqrt{2\pi\sigma^2}} \int_{a_k}^{e_{k+1}} dx\, e^{-\frac{(x-E_\beta)^2}{2\sigma^2}} \geq \tilde{\delta}_k \frac{1}{\sqrt{2\pi\sigma^2}} \exp\left(-\max_{x \in [a_k, e_{k+1}]} \frac{(x-E_\beta)^2}{2\sigma^2}\right). \tag{27}$$

Hence, in order for $G(e_{k+1}) - G(a_k) \geq 3\tilde{\zeta}_N$ we need to pick $a_k$ such that

$$\tilde{\delta}_k \geq 3\sqrt{2\pi}\sigma\tilde{\zeta}_N \exp\left(\max_{x \in [a_k, e_{k+1}]} \frac{(x - E_\beta)^2}{2\sigma^2}\right). \tag{28}$$

We have

$$\frac{|x - E_\beta|}{\sqrt{2}\sigma} \leq \frac{|x - E| + |E - E_\beta|}{\sqrt{2}\sigma} \leq \frac{1}{\sqrt{2}}\left(2 + \frac{\Delta(k)}{\sigma}\right) \tag{29}$$

where we have used that by assumption $|E - E_\beta| \leq \sigma$ and that since $x \in [a_k, e_{k+1}]$ we have

$$|x - E| \leq \Delta(k) + \delta_k \leq \Delta(k) + \sigma, \tag{30}$$

where we have imposed $\delta_k \leq \sigma$. We set

$$\tilde{\delta}_k = \delta_k^* := 3\sqrt{2\pi}\sigma\tilde{\zeta}_N \exp\left(\frac{1}{2}\left(\frac{\Delta(k)}{\sigma} + 2\right)^2\right). \tag{31}$$

We have $G(e_{k+1}) - G(a_k) \geq 3\tilde{\zeta}_N$, and naturally, since $\tilde{\delta}_k \leq \delta_k$, in order to be able to choose $\tilde{\delta}_k = \delta_k^*$, we need $\delta_k \geq \delta_k^*$. Hence if

$$\delta_k^* \leq \delta_k \leq \sigma \tag{32}$$

we have

$$\frac{Z(\beta)}{Z_k(\beta)} \leq \frac{1}{\tilde{\zeta}_N}. \tag{33}$$

We need to ensure that the ranges for $\delta_k$ exist. Notice that since by definition $\Delta(k) \leq \Delta$

$$\frac{\Delta(k)}{\sigma} + 2 \leq \sqrt{\ln\left(\frac{\epsilon N^{\frac{1}{D+1}}}{R \ln(N)^{2D}}\right)} + 2 \leq \sqrt{2\ln\left(\frac{\epsilon N^{\frac{1}{D+1}}}{R \ln(N)^{2D}}\right)}, \tag{34}$$

where we for the last inequality we have used

$$\epsilon \geq e^{1+\sqrt{5}} \ln(N)^{2D} N^{\frac{-1}{D+1}}. \tag{35}$$

Hence,

$$\delta_k^* \leq 3\sqrt{2\pi}\epsilon N^{\frac{1}{D+1}} \frac{\sigma\tilde{\zeta}_N}{R\ln(N)^{2D}} \leq \frac{1}{2}\sigma \tag{36}$$

where we have used the assumed upper bound on $\epsilon$. Hence if for some $k$ we have $\Delta(k) + \delta_k \geq \Delta$, that is, the lower bound on the next interval requires to leave $I$, we can simply redefine $\delta_{k-1}$ to extend all the way to the edge of $I$. We can now go back to equation Eq. (23) and we have

$$S(\tau \| g_\beta(H)) \leq -S(\tau) + \sum_k \text{tr}(\tau_k) \log(d_k) + \beta \log(e) 3\sqrt{2\pi}\sigma\tilde{\zeta}_N \epsilon N^{\frac{1}{D+1}} + \log\left(\frac{1}{\tilde{\zeta}_N}\right). \tag{37}$$

Furthermore, we use

$$\epsilon \geq \log\left(\frac{1}{\tilde{\zeta}_N}\right) N^{\frac{-1}{D+1}} \tag{38}$$

which ensures that $\log\left(1/\tilde{\zeta}_N\right) \leq \epsilon N^{\frac{1}{D+1}}$, hence overall,

$$S(\tau \| g_\beta(H)) \leq -S(\tau) + \sum_k \text{tr}(\tau_k) \log(d_k) + \left(\beta \log(e) 3\sqrt{2\pi}\sigma\tilde{\zeta}_N + 1\right) \epsilon N^{\frac{1}{D+1}}. \tag{39}$$

If we impose

$$\sum_k \mathrm{tr}\,(\tau_k) \log(d_k) - S(\tau) \leq \epsilon N^{\frac{1}{D+1}}, \tag{40}$$

we have that

$$S(\tau || g_\beta(H)) \leq (\beta \log(e) 3\sqrt{2\pi}\sigma\tilde{\zeta}_N + 2)\epsilon N^{\frac{1}{D+1}} \leq (\beta \log(e) 3\sqrt{2\pi} + 2)\epsilon N^{\frac{1}{D+1}}, \tag{41}$$

where we have used $\sigma\tilde{\zeta}_N \leq R \ln(N)^{2D}$. The result then follows from Corollary 1. $\square$

Theorem 1 and its approximate version (Eq. (11) in the main text) follow from the following more general statement.

**Theorem 2** (Ensemble equivalence). *Let $H$ be a local Hamiltonian and $\beta$ be an inverse temperature for which the Gibbs state $g_\beta(H)$ has exponential decay of correlations and standard deviation $\sigma \geq \Omega(\sqrt{N})$. Suppose $\sigma\zeta_N \leq R \ln(N)^{2D} N^{-\kappa}$ for $R, \kappa \geq 0$ constant. Let $\omega$ denote an approximate GmE state with $\Delta, \delta, \eta$ satisfying*

$$\begin{aligned} e^{\Delta^2/\sigma^2} &\leq \frac{N^{\frac{1-\alpha}{D+1}}}{R \ln(N)^{2D}}, \\ 3\sqrt{2\pi} N^{\frac{1-\alpha}{D+1}-\kappa} &\leq \delta \leq \sigma, \\ \eta &\leq N^{\frac{1-\alpha}{D+1}} \end{aligned} \tag{42}$$

*with $\alpha \in [0, 1)$, such that $|E - E_\beta| \leq \sigma$. Then for any side length $l$ such that $l^D \leq C_1 N^{\frac{1}{D+1}-\gamma_1\alpha}$, the following holds*

$$D_l(\omega, g_\beta(H)) \leq C_2 N^{-\gamma_2\alpha} + 2(1-p_\Delta), \tag{43}$$

*with $C_1, C_2$ being system-size independent constants, and $\gamma_1, \gamma_2$ only depend on the dimension of the lattice $D$.*

*Proof.* Choose

$$\epsilon := N^{-\frac{\alpha}{D+1}} \tag{44}$$

with $\alpha < 1$ constant and set $\tilde{\zeta}_N := R \ln(N)^{2D} N^{-\kappa}/\sigma \geq \zeta_N$. Then we have

$$\Delta \leq \sigma \sqrt{\ln\left(\frac{N^{\frac{1-\alpha}{D+1}}}{R \ln(N)^{2D}}\right)} = \sigma \sqrt{\ln\left(\frac{\epsilon N^{\frac{1}{D+1}}}{R \ln(N)^{2D}}\right)}. \tag{45}$$

We would like to apply Theorem 1. For this, we set $\delta_k := \delta$, hence we need $\delta \geq \delta_k^*$ for all $k$. As shown in the proof of Theorem 1, for $N$ sufficiently large we have by using $\Delta(k) \leq \Delta$

$$\delta_k^* \leq 3\sqrt{2\pi}\epsilon N^{\frac{1}{D+1}} \frac{\sigma\tilde{\zeta}_N}{R \ln(N)^{2D}} = 3\sqrt{2\pi} N^{\frac{1-\alpha}{D+1}-\kappa} \tag{46}$$

Then, for all $k$, choose

$$\delta = 3\sqrt{2\pi} N^{\frac{1-\alpha}{D+1}-\kappa}. \tag{47}$$

The subdivision then satisfies the assumptions of Theorem 1 for the chosen $\epsilon$. Let $\tau = \sum_{k=1}^K q_k \tilde{\omega}_{\delta_k}$ such that $\omega = p_\Delta \tau + (1-p_\Delta)\rho_{\mathrm{tail}}$. We have

$$\|\omega - \tau\|_1 = \|\omega - p_\Delta \tau + p_\Delta \tau - \tau\|_1 \leq 2(1-p_\Delta). \tag{48}$$

We then have

$$D_l(\omega, g_\beta(H)) \leq D_l(\tau, g_\beta(H)) + 2(1-p_\Delta) \tag{49}$$

Furthermore, by the concavity of the Von Neumann entropy

$$S(\tau) \geq \sum_k q_k S(\tilde{\omega}_{\delta_k}) \geq \sum_k q_k S(\omega_{\delta_k}) - N^{\frac{1-\alpha}{D+1}}. \tag{50}$$

Notice that $\mathrm{tr}\,(\Pi_k \tau) = q_k$ and $S(\omega_{\delta_k}) = \log(d_k)$. By Theorem 1,

$$D_l(\tau, g_\beta(H)) \leq C_2 N^{-\gamma_2 \alpha}, \tag{51}$$

for $C_2, \gamma_2$ constants. $\square$

Theorem 1 in the main text is the special case where $p_\Delta = 1$ and $\tilde{\omega}_{\delta_k} = \omega_{\delta_k}$.

## II. WEINGARTEN CALCULUS

In what follows, we will often need to compute averages of random unitaries. Here we very briefly give the necessary basics and notation. For a comprehensive introduction, see [3]. Let $\mathcal{U}(d)$ be the unitary group of dimension $d$ and $\mu_H(d)$ be the Haar measure thereon. We will be interested in computing averages involving unitaries of the form $U^{\otimes k}$, for some integer $k$, where $U$ is drawn from $\mu_H(d)$. In reality, $U$ needs only to be drawn from a unitary $k$-design. In particular, we have, for any operator $X$ acting on $\mathbb{C}^{d^{\otimes k}}$,

$$\mathbb{E}_{U\sim\mu_H(d)}\left(U^{\otimes k} X U^{\otimes k\dagger}\right) = \sum_{\sigma,\tau\in S_k} \mathrm{Wg}(\sigma\tau^{-1},d)\mathrm{tr}\left(R_\sigma X\right) R_\tau \tag{52}$$

where $S_k$ is the symmetric group of $k$ elements with $R$ being a representation such that for $\sigma \in S_k$, $R_\sigma$ permutes the tensor spaces

$$R_\sigma |i_1,\ldots,i_k\rangle = |i_{\sigma(1)},\ldots,i_{\sigma(k)}\rangle \tag{53}$$

in $\mathbb{C}^{d^{\otimes k}}$ according to $\sigma$. $\mathrm{Wg}(\sigma\tau^{-1},d)$ is the Weingarten function. Table I and Table II give its values for $k=2$ and $k=4$, respectively. Such tables can be generated using the software developed in [4] and tables up to $k=20$ are available at the corresponding GitHub repository.

| Cycle type of $\sigma$ | $\mathrm{Wg}(\sigma,d)$ |
|---|---|
| $(1,1)$ | $\frac{1}{d^2-1}$ |
| $(2)$ | $\frac{1}{-d(d^2-1)}$ |

TABLE I. Weingarten functions for $k=2$.

| Cycle type of $\sigma$ | $\mathrm{Wg}(\sigma,d)$ |
|---|---|
| $(1,1,1,1)$ | $\frac{d^4-8d^2+6}{d^8-14d^6+49d^4-36d^2}$ |
| $(2,1,1)$ | $\frac{1}{d^5-10d^3+9d}$ |
| $(3,1)$ | $\frac{2d^2-3}{d^8-14d^6+49d^4-36d^2}$ |
| $(2,2)$ | $\frac{d^2+6}{d^8-14d^6+49d^4-36d^2}$ |
| $(4)$ | $-\frac{1}{d^7-14d^5+49d^3-36d}$ |

TABLE II. Weingarten functions for $k=4$.

## III. THERMALIZATION UNDER TYPICAL HAMILTONIANS

We now apply our thermality condition to prove thermalization with high probability of appropriate states under a suitable slightly perturbed Hamiltonian. Before proving the theorems in the main text, we discuss tails decay for states with exponential decay of correlations. The following has been proven in ref. [5].

**Theorem 3** (Spectral tail bound). *Let $\rho$ be a state and $H$ be a local Hamiltonian. Let $\Pi_\Delta$ be the projector onto all eigenstates of $H$ with energy $E$ such that $|E - \mathrm{tr}(\rho H)| \leq \Delta$. Then,*

1. *if $\rho$ is a product state, and $\Delta \geq g_0\sqrt{N}$*

$$\mathrm{tr}\left((1-\Pi_\Delta)\rho\right) \leq \mathrm{e}^{-s_0\frac{\Delta^2}{N}}, \tag{54}$$

*where $g_0, s_0$ are system size independent constants.*

2. if $\rho$ has exponential decay of correlations, and $\Delta \geq g_\xi \sqrt{N}$

$$\operatorname{tr}\left((1 - \Pi_\Delta)\rho\right) \leq \mathrm{e}^{-s_\xi \left(\frac{\Delta^2}{N}\right)^{\frac{1}{D+1}}}, \tag{55}$$

where $g_\xi, s_\xi$ are system size independent constants depending on the correlation length $\xi$.

We repeat here the definition of the ensemble $\mathcal{E}(\delta)$: we divide the energy spectrum into intervals $\{I_k\}_{k=1}^K$ of equal energy width $\delta$, the eigenstates in the $k$-th window span a vector space which we call $\mathcal{W}_k$. $\mathcal{E}(\delta)$ is then the ensemble of random unitaries of the form

$$U = \bigoplus_{k=1}^K U_k, \tag{56}$$

where $U_k$ is drawn from the Haar measure on the unitary group acting on $\mathcal{W}_k$ independently for every $k$. Generally, we will denote by $\Pi_k$ the projector onto $\mathcal{W}_k$, by $W_k$ the set of indices $\nu$ such that $E_\nu \in I_k$, and by $d_k := |W_k|$ the number of eigenstates contained in each window. In addition we will denote $\rho_k = \Pi_k \rho \Pi_k$. For convenience, let us prove a fact that will be useful later.

**Lemma 3** (Perturbing Hamiltonians). *Let $U$ be drawn from $\mathcal{E}(\delta)$. Then*

$$\|H - UHU^\dagger\|_\infty \leq \delta. \tag{57}$$

*Proof.* Since both $H$ and $U$ commute with $\Pi_k$, we have

$$(UHU^\dagger - H)^2 = \sum_{k=1}^K \Pi_k (UHU^\dagger - H)^2 \Pi_k = \sum_{k=1}^K (U_k H_k U_k^\dagger - H_k)^2 \tag{58}$$

with $H_k := \Pi_k H \Pi_k$. Let $\rho$ be a state and define $\rho_k := \Pi_k \rho \Pi_k$. We then have

$$\operatorname{tr}\left(\rho(UHU^\dagger - H)^2\right) = \sum_{k=1}^K \operatorname{tr}\left(\rho_k (U_k H_k U_k^\dagger - H_k)^2\right) \leq \sum_{k=1}^K \operatorname{tr}(\rho_k) \|U_k H_k U_k^\dagger - H_k\|_\infty^2, \tag{59}$$

where we have used Hölder's inequality and the fact that since $\rho_k > 0$, $\|\rho_k\|_1 = \operatorname{tr}(\rho_k)$. Define

$$m_k := e_k + \frac{\delta}{2} = \frac{e_k + e_{k+1}}{2}, \tag{60}$$

then

$$\|U_k H_k U_k^\dagger - H_k\|_\infty = \|U_k H_k U_k^\dagger - m_k \Pi_k + m_k \Pi_k - H_k\|_\infty \leq 2\|H_k - m_k \Pi_k\|_\infty \leq \delta^2. \tag{61}$$

Hence,

$$\operatorname{tr}\left(\rho(UHU^\dagger - H)^2\right) \leq \delta^2, \tag{62}$$

which shows $\|UHU^\dagger - H\|_\infty \leq \delta$ by the definition of operator norm. $\square$

First, we prove Lemma 1 in the main text in the more general case where eigenvalues can be degenerate.

**Lemma 4** (Approximate GmE at equilibrium). *Let $\rho$ be a product state and $H$ be a local Hamiltonian. Let*

$$\rho_\infty^{UHU^\dagger} = \lim_{T \to \infty} \frac{1}{T} \int_0^T \mathrm{e}^{-iUHU^\dagger t} \rho \, \mathrm{e}^{iUHU^\dagger t} \tag{63}$$

*where $U$ is drawn from $\mathcal{E}(\delta)$. Consider the interval $I = [E - \Delta, E + \Delta]$ around $E = \operatorname{tr}(\rho H)$ with $\Delta \geq \omega(\sqrt{N})$ an integer multiple of $\delta$, then*

$$\rho_\infty^{UHU^\dagger} = p_\Delta \left(\sum_{k: I_k \subset I} q_k \tilde{\omega}_{\delta_k}\right) + (1 - p_\Delta)\rho_{\mathrm{tail}} \tag{64}$$

*with $p_\Delta \geq 1 - \mathrm{e}^{-c_1 \frac{\Delta^2}{N}}$ and $\sum_{k: I_k \subset I} q_k = 1$. Moreover, with probability at least $1 - (\mathcal{D}+1)2^{-r}$, where $\mathcal{D}$ is the degeneracy of the most degenerate eigenvalue with energy inside $I$, we have*

$$\sum_{k: I_k \subset I} q_k (S(\omega_{\delta_k}) - S(\tilde{\omega}_{\delta_k})) \leq r, \tag{65}$$

*where $c_1$ is a system-size independent constant.*

*Proof.* Remember that we denote the spectral decomposition of $H$ as $H = \sum_\nu E_\nu P_\nu$. We have

$$\rho_\infty^{UHU^\dagger} = \sum_\nu U P_\nu U^\dagger \rho U P_\nu U^\dagger. \tag{66}$$

Notice that if $E_\nu \in I_k$, then $P_\nu$ is a projector in $\mathcal{W}_k$, and since the unitaries $U$ preserve the spaces $\mathcal{W}_k$, $UP_\nu U^\dagger$ is also a projector in $\mathcal{W}_k$. Let $W_k$ be the set of indices $\nu$ such that $E_\nu \in I_k$. Then

$$\rho_\infty^{UHU^\dagger} = \sum_{k: I_k \subset I} \sum_{\nu \in W_k} U_k P_\nu U_k^\dagger \rho_k U_k P_\nu U_k^\dagger + \sum_{k: I_k \not\subset I} \sum_{\nu \in W_k} U_k P_\nu U_k^\dagger \rho_k U_k P_\nu U_k^\dagger =: p_\Delta \underbrace{\sum_{k: I_k \subset I} \frac{p_k}{p_\Delta}}_{=:q_k} \tilde{\omega}_{\delta_k} + (1-p_\Delta)\rho_{\text{tail}} \tag{67}$$

with

$$p_\Delta = \operatorname{tr}\left(\Pi_\Delta \rho_\infty^{UHU^\dagger}\right), \tag{68}$$

$$p_k = \sum_{\nu \in W_k} \operatorname{tr}\left(\rho U_k P_\nu U_k^\dagger\right) = \operatorname{tr}(\rho \Pi_k), \tag{69}$$

$$\tilde{\omega}_{\delta_k} = \frac{1}{p_k} \sum_{\nu \in W_k} U_k P_\nu U_k^\dagger \rho_k U_k P_\nu U_k^\dagger, \tag{70}$$

$$\rho_{\text{tail}} = \frac{1}{1-p_\Delta} \sum_{k: I_k \not\subset I} \sum_{\nu \in W_k} U_k P_\nu U_k^\dagger \rho_k U_k P_\nu U_k^\dagger. \tag{71}$$

To proceed, we start by bounding the entropy. We have

$$S(\tilde{\omega}_{\delta_k}) \geq S_2(\tilde{\omega}_{\delta_k}) = -\log\left(\operatorname{tr}\left(\tilde{\omega}_{\delta_k}^2\right)\right) \tag{72}$$

and

$$\operatorname{tr}\left(\tilde{\omega}_{\delta_k}^2\right) = \frac{1}{p_k^2} \sum_{\nu \in W_k} \operatorname{tr}\left((U_k P_\nu U_k^\dagger \rho_k)^2\right) = \frac{1}{p_k^2} \sum_{\nu \in W_k} \operatorname{tr}\left(U_k^{\otimes 2} P_\nu^{\otimes 2} U_k^{\otimes 2 \dagger} \rho^{\otimes 2} \mathbb{F}\right) \tag{73}$$

where $\mathbb{F} = R_{(1,2)}$ flips the two tensor spaces. Then using the Weingarten formula as outlined in Section II yields

$$\mathbb{E}_{U \sim \mathcal{E}(\delta)}\left(\operatorname{tr}\left(\tilde{\omega}_{\delta_k}^2\right)\right) = \frac{1}{p_k^2} \sum_{\nu \in W_k} \frac{\operatorname{tr}(P_\nu)}{d_k^2 - 1}\left[\operatorname{tr}\left((\rho_k)^2\right)\left(\operatorname{tr}(P_\nu) - \frac{1}{d_k}\right) + \operatorname{tr}(\rho_k)^2\left(1 - \frac{\operatorname{tr}(P_\nu)}{d_k}\right)\right]$$
$$\leq \frac{\operatorname{tr}(\rho_k)^2}{p_k^2} \sum_{\nu \in W_k} \frac{\operatorname{tr}(P_\nu)(1 + \operatorname{tr}(P_\nu))}{d_k(d_k+1)} \leq \frac{1}{d_k+1}(1 + \max_\nu \operatorname{tr}(P_\nu)) =: \frac{D_k+1}{d_k+1}, \tag{74}$$

where $D_k$ is the degeneracy of the maximally degenerate eigenvalue inside $W_k$. By Markov's inequality

$$\Pr_{U \sim \mathcal{E}(\delta)}\left[\log(d_k) - S(\tilde{\omega}_{\delta_k}) \geq r\right] \leq \Pr_{U \sim \mathcal{E}(\delta)}\left[\operatorname{tr}\left(\tilde{\omega}_{\delta_k}^2\right) \geq \frac{1}{d_k} 2^r\right] \leq \frac{(D_k+1)d_k}{d_k+1} 2^{-r} \leq (D_k+1)2^{-r}. \tag{75}$$

Noticing that $\log(d_k) = S(\omega_{\delta_k})$ yields the result. Finally, we need to bound $p_\Delta$. Notice that

$$p_\Delta = \operatorname{tr}\left(\Pi_\Delta \rho_\infty^{UHU^\dagger}\right) = \sum_{k: I_k \subset I} \sum_{\nu \in W_k} \operatorname{tr}\left(\rho U P_\nu U^\dagger\right) = \operatorname{tr}(\Pi_\Delta \rho). \tag{76}$$

We now wish to bound $p_\Delta$ using Theorem 3. By construction if $E_\nu \notin I$, then

$$|\operatorname{tr}(\rho H) - E_\nu| \geq \Delta \geq \omega(\sqrt{N}). \tag{77}$$

Then, by Theorem 3, we have

$$p_\Delta \geq 1 - e^{-g_0 \frac{\Delta^2}{N}} \tag{78}$$

if $\rho$ is a product state and

$$p_\Delta \geq 1 - e^{-g_\xi \left(\frac{\Delta^2}{N}\right)^{\frac{1}{D+1}}} \tag{79}$$

if $\rho$ has exponential decay of correlations. $\square$

Now the following version of Theorem 2 in the main text follows from Theorem 2 and Lemma 4.

**Theorem 4** (Typical thermalization). *Let $H$ be a $k$-local Hamiltonian and $\rho$ be a product state. Let $g_\beta(H)$ be the Gibbs state of $H$ at inverse temperature $\beta$ such that $|\mathrm{tr}\,(g_\beta(H)H) - \mathrm{tr}\,(\rho H)| \leq \sigma$. Assume $g_\beta(H)$ has exponential decay of correlations and $\sigma\zeta_N \leq R\ln(N)^{2D}N^{-\kappa}$ for $B, \kappa \geq 0$ constants. For any constant $\alpha \in [0,1)$, if $\delta = 3\sqrt{2\pi}N^{\frac{1-\alpha}{D+1}-\kappa}$, then with probability at least $1 - (\mathcal{D}+1)\exp(-c_2 N^{\frac{1-\alpha}{D+1}})$ drawing $U$ at random from $\mathcal{E}(\delta)$, we have*

$$D_l(\rho_\infty^{UHU^\dagger}, g_\beta(H)) \leq C_2 \, N^{-\gamma_2\alpha} + N^{-\gamma_3(1-\alpha)}\ln(N)^{2D}, \tag{80}$$

*where $c_2, C_2, \gamma_2, \gamma_3$ are system-size independent constants.*

*Proof.* Let $\Delta$ such that

$$e^{\Delta^2/\sigma^2} = \frac{N^{\frac{1-\alpha}{D+1}}}{R\ln(N)^{2D}}. \tag{81}$$

By applying Lemma 4 with $r := N^{\frac{1-\alpha}{D+1}}$ we have that

$$\rho_\infty^{UHU^\dagger} = p_\Delta \tau + (1-p_\Delta)\rho_{\mathrm{tail}} \tag{82}$$

is approximately GmE satisfying the conditions of Theorem 2, hence we have

$$D_l(\rho_\infty^{UHU^\dagger}, g_\beta(H)) \leq C_2 \, N^{-\gamma_2\alpha} + 2(1-p_\Delta) \leq C_2 \, N^{-\gamma_2\alpha} + N^{-\gamma_3(1-\alpha)}\ln(N)^{2D}. \tag{83}$$

□

A similar theorem is immediate for states with exponential decay of correlations, but in this case

$$1 - p_\Delta \leq \exp\left(-\Omega\left(\ln\left(\frac{N^{\frac{1-\alpha}{D+1}}}{\ln(N)^{2D}}\right)\right)^{\frac{1}{D+1}}\right). \tag{84}$$

Equilibration also immediately follows from Lemma 4, as a matter of fact for any observable $A$ in the case of non-degenerate energy gaps, that is if $E_\mu - E_\nu = E_{\mu'} - E_{\nu'}$ implies $(\mu,\nu) = (\mu',\nu')$. We have [6]

$$\Delta A_\infty := \lim_{T\to\infty} \frac{1}{T}\int_0^T \mathrm{d}t\,\mathrm{tr}\left(A(\rho^{UHU^\dagger}(t) - \rho_\infty^{UHU^\dagger})\right)^2 \leq \|A\|_\infty^2 \mathrm{tr}\left(\left(\rho_\infty^{UHU^\dagger}\right)^2\right) \leq \|A\|_\infty^2\left((1-p_\Delta)^2 + \sum_{k:I_k\subset I} q_k^2 \mathrm{tr}\,(\tilde{\omega}_{\delta_k}^2)\right)$$

$$\leq \|A\|_\infty^2\left((1-p_\Delta)^2 + 2^r \sum_{k:I_k\subset I} q_k^2 \frac{1}{d_k}\right) \leq \|A\|_\infty^2\left((1-p_\Delta)^2 + 2^r \max_k \frac{1}{d_k}\right). \tag{85}$$

The condition of non-degenerate gaps is expected to hold generically for interacting systems. Then for example taking $r = \min_k \log(\sqrt{d_k})$ we have that with probability at least $1 - (\mathcal{D}+1)\max_k \frac{1}{\sqrt{d_k}}$, $\Delta A_\infty \leq \|A\|_\infty^2\left((1-p_\Delta)^2 + 2^r \max_k \frac{1}{\sqrt{d_k}}\right)$. Since $d_k$ is expected to be exponentially large, the system equilibrates. In addition, we have the following statement.

**Theorem 5** (Short-time evolution). *For any state $\rho$ and $U \sim \mathcal{E}(\delta)$ and any $t \geq 0$*

$$\|e^{-iHt}\rho e^{iHt} - e^{-iUHU^\dagger t}\rho e^{iUHU^\dagger t}\|_1 \leq 2t\delta. \tag{86}$$

*Proof.* Let $V = e^{-iHt}$ and $\tilde{V} = e^{-iUHU^\dagger t}$. Then

$$\|V\rho V^\dagger - \tilde{V}\rho\tilde{V}^\dagger\|_1 = \|\tilde{V}^\dagger V\rho V^\dagger\tilde{V} - \rho\|_1 = \|\tilde{V}^\dagger V[\rho, V^\dagger\tilde{V}]\|_1 = \|[\rho, V^\dagger\tilde{V} - \mathbb{1}]\|_1 \leq 2\|\rho\|_1\|V - \tilde{V}\|_\infty \tag{87}$$

where we have used Hölder's inequality and standard properties of the trace norm. Let $X = UHU^\dagger - H$, then

$$\|V - \tilde{V}\|_\infty = \|e^{iHt} - e^{iHt+iXt}\|_\infty. \tag{88}$$

Now let the function $f$ be $f(s) = e^{iHt+iXts}$, then

$$\|V - \tilde{V}\|_\infty = \|f(1) - f(0)\|_\infty \leq \int_0^1 \mathrm{d}s\|f'(s)\|_\infty. \tag{89}$$

Then, using

$$f'(s) = it \int_0^1 dr \, e^{r(iHt+iXts)} X e^{(1-r)(iHt+iXts)}, \tag{90}$$

we get

$$\|f'(s)\|_\infty \leq t\|X\|_\infty \tag{91}$$

and hence by Lemma 3

$$\|V - \tilde{V}\|_\infty \leq t\|X\|_\infty = t\|H - UHU^\dagger\|_\infty \leq t\delta, \tag{92}$$

together with $\|\rho\|_1 = 1$ this proves the result. $\square$

Choosing

$$\delta = \Omega(N^{\frac{1-\alpha}{D+1} - \kappa}), \tag{93}$$

if the Berry-Esseen error is bounded by $\tilde{O}(N^{-1/2-\kappa})$, we have, again choosing $\alpha \geq 1 - \kappa(D+1)$, that $\delta$ is a decreasing function of the system size, and the two dynamics are indistinguishable up to $\epsilon$ until a time $\sim \epsilon N^{\kappa - \frac{1-\alpha}{D+1}}$. We now move on to proving that the Gibbs states of $H$ and $UHU^\dagger$ are locally indistinguishable.

**Theorem 6** (Local indistinguishability of perturbed Gibbs states). *Let $H$ be a $k$-local Hamiltonian. Let $g_\beta(H)$ be the Gibbs state of $H$ at an inverse temperature $\beta$ such that $g_\beta(H)$ has exponential decay of correlations. For any $U$ drawn from $\mathcal{E}(\delta)$ and $l, \kappa, \alpha$ as in Theorem 2 we have*

$$D_l(g_\beta(H), g_\beta(UHU^\dagger)) \leq C_3 \, N^{-\gamma_4 \alpha - \gamma_5 \kappa}, \tag{94}$$

*for system-size independent constants $C_3, \gamma_4, \gamma_5$.*

*Proof.* A simple computation reveals

$$S(g_\beta(UHU^\dagger)\|g_\beta(H)) = \beta \text{tr}\left((UHU^\dagger - H)g_\beta(H)\right) \leq \beta\|H - UHU^\dagger\|_\infty \leq \beta\delta \tag{95}$$

where we have used Lemma 3. The result follows from Corollary 1, with $\epsilon = \beta\delta N^{-\frac{1}{D+1}} = \beta N^{-\frac{\alpha}{D+1} - \kappa}$, which ends the proof. $\square$

## IV. RELAXATION DYNAMICS

We previously discussed the dynamics in the short-time regime, we now turn to discussing the typical late-time relaxation dynamics to the thermal state following an non-equilibrium initial preparation. We abbreviate in the following $\rho_U(t) := e^{-iUHU^\dagger t}\rho e^{iUHU^\dagger t}$. To prove the statement in the main text, we write

$$\langle A \rangle_{\rho_U(t)} - \langle A \rangle_{\rho_\infty^{UHU^\dagger}} = \langle A \rangle_{\rho_U(t)} - \mathbb{E}\left(\langle A \rangle_{\rho_U(t)}\right) + \mathbb{E}\left(\langle A \rangle_{\rho_U(t)}\right) - \mathbb{E}\left(\langle A \rangle_{\rho_\infty^{UHU^\dagger}}\right) + \mathbb{E}\left(\langle A \rangle_{\rho_\infty^{UHU^\dagger}}\right) - \langle A \rangle_{\rho_\infty^{UHU^\dagger}} \tag{96}$$

where the expectation values are taken over $U \sim \mathcal{E}(\delta)$. Hence

$$\left|\langle A \rangle_{\rho_U(t)} - \langle A \rangle_{\rho_\infty^{UHU^\dagger}}\right| \leq \left|\langle A \rangle_{\rho_U(t)} - \mathbb{E}\left(\langle A \rangle_{\rho_U(t)}\right)\right| + \left|\langle A \rangle_{\rho_\infty^{UHU^\dagger}} - \mathbb{E}\left(\langle A \rangle_{\rho_\infty^{UHU^\dagger}}\right)\right| + R(t), \tag{97}$$

where we have defined the function $R$ to be

$$R(t) := \left|\mathbb{E}\left(\langle A \rangle_{\rho_U(t)}\right) - \mathbb{E}\left(\langle A \rangle_{\rho_\infty^{UHU^\dagger}}\right)\right|. \tag{98}$$

We now need to show that the first two terms are small with high probability, and the promised decay of $R(t)$ with the relevant spectral assumptions.

In order to show that the first two terms are small with high probability, we will need to bound their second moments. From now on, let $\Pi_k$ be the orthogonal projector onto the $k$-th window $\mathcal{W}_k$, and $d_k = \text{tr}(\Pi_k)$ the number of energy eigenstates contained within. We denote as $K$ the total number of windows. In addition, for any operator $X$, we define

$$X_{k,j} := \Pi_k X \Pi_j \qquad X_k = X_{k,k}. \tag{99}$$

We also define

$$\phi_k(t) := \frac{1}{d_k}\mathrm{tr}\left(\Pi_k e^{-iHt}\right). \tag{100}$$

We now state the second moment bounds, which are proven in Section IV A under the assumption that $H$ has no degenerate eigenstates. If $H$ has weak degeneracies we expect very similar results to hold.

**Lemma 5.**

$$\mathbb{E}\left(\left|\langle A\rangle_{\rho_U(t)} - \mathbb{E}\left(\langle A\rangle_{\rho_U(t)}\right)\right|^2\right) \leq \|A\|_\infty^2 O\left(K^2 \sum_k \mathrm{tr}\,(\rho_k)\frac{1}{d_k}\right). \tag{101}$$

**Lemma 6.**

$$\mathbb{E}\left(\left|\langle A\rangle_{\rho_\infty^{UHU^\dagger}} - \mathbb{E}\left(\langle A\rangle_{\rho_\infty^{UHU^\dagger}}\right)\right|^2\right) \leq \|A\|_\infty^2 O\left(\sum_k \mathrm{tr}\,(\rho_k)\frac{1}{d_k}\right). \tag{102}$$

These bounds imply the following:

**Lemma 7** (Time-evolution concentration). *For a Hamiltonian $H$, an observable $A$, and some initial state $\rho$ it holds that*

$$\Pr_{U\sim\mathcal{E}(\delta)}\left[\left|\langle A\rangle_{\rho_U(t)} - \mathbb{E}\left(\langle A\rangle_{\rho_U(t)}\right)\right| + \left|\langle A\rangle_{\rho_\infty^{UHU^\dagger}} - \mathbb{E}\left(\langle A\rangle_{\rho_\infty^{UHU^\dagger}}\right)\right| \geq \epsilon\right] \leq \|A\|_\infty^2 \frac{1}{\epsilon^2} O\left(K^2 \sum_k \mathrm{tr}\,(\rho_k)\frac{1}{d_k}\right). \tag{103}$$

$\sum_k \mathrm{tr}\,(\rho_k)\frac{1}{d_k}$ is the average number of eigenstates in the windows where $\rho_k$ has a non negligible support. Since these windows are all in the bulk of the spectrum, it is expected that $\sum_k \mathrm{tr}\,(\rho_k)\frac{1}{d_k} \leq e^{-\Omega(N)}$.

*Proof.* Via a union bound, we get

$$\Pr_{U\sim\mathcal{E}(\delta)}\left[\left|\langle A\rangle_{\rho_U(t)} - \mathbb{E}\left(\langle A\rangle_{\rho_U(t)}\right)\right| + \left|\langle A\rangle_{\rho_\infty^{UHU^\dagger}} - \mathbb{E}\left(\langle A\rangle_{\rho_\infty^{UHU^\dagger}}\right)\right| \geq \epsilon\right]$$
$$\leq \Pr_{U\sim\mathcal{E}(\delta)}\left[\left|\langle A\rangle_{\rho_\infty^{UHU^\dagger}} - \mathbb{E}\left(\langle A\rangle_{\rho_\infty^{UHU^\dagger}}\right)\right| \geq \epsilon/2\right] + \Pr_{U\sim\mathcal{E}(\delta)}\left[\left|\langle A\rangle_{\rho_U(t)} - \mathbb{E}\left(\langle A\rangle_{\rho_U(t)}\right)\right| \geq \epsilon/2\right]$$
$$\leq \mathbb{E}\left(\left|\langle A\rangle_{\rho_U(t)} - \mathbb{E}\left(\langle A\rangle_{\rho_U(t)}\right)\right|^2\right)\frac{4}{\epsilon^2} + \mathbb{E}\left(\left|\langle A\rangle_{\rho_\infty^{UHU^\dagger}} - \mathbb{E}\left(\langle A\rangle_{\rho_\infty^{UHU^\dagger}}\right)\right|^2\right)\frac{4}{\epsilon^2} \leq \|A\|_\infty^2\frac{1}{\epsilon^2} O\left(K^2 \sum_k \mathrm{tr}\,(\rho_k)\frac{1}{d_k}\right). \tag{104}$$

$\square$

We now move on to bounding $R(t)$. We have

$$\langle A\rangle_{\rho_U(t)} = \sum_{k,j} \mathrm{tr}\left(A_{j,k} U_k e^{-iH_k t} U_k^\dagger \rho_{k,j} U_j e^{iH_j t} U_j^\dagger\right) \tag{105}$$

and

$$\langle A\rangle_{\rho_\infty^{UHU^\dagger}} = \sum_k \sum_{\nu\in W_k} \mathrm{tr}\left(P_\nu U_k^\dagger \rho_k U_k P_\nu U_k^\dagger A U_k\right) \tag{106}$$

where $W_k$ is the set of all $\nu$ such that $E_\nu \in I_k$. Invoking again the Weingarten formula as outlined in Section II, simple computations reveal for the expectation value

$$\mathbb{E}\left(\langle A\rangle_{\rho_U(t)}\right) = \sum_{k\neq j}\phi_k(t)\phi_j^*(t)\mathrm{tr}\,(A_{j,k}\rho_{k,j}) + \sum_k \frac{|\phi_k(t)|^2 d_k^2 - 1}{d_k^2 - 1}\mathrm{tr}\,(A_k\rho_k) + \frac{d_k}{d_k^2-1}(1-|\phi_k(t)|^2)\mathrm{tr}\,(A_k)\,\mathrm{tr}\,(\rho_k), \tag{107}$$

as well as

$$\mathbb{E}\left(\langle A\rangle_{\rho_\infty^{UHU^\dagger}}\right) = \sum_k \frac{|\overline{\phi}_k|^2 d_k^2 - 1}{d_k^2 - 1}\mathrm{tr}\,(A_k\rho_k) + \frac{d_k}{d_k^2-1}(1-|\overline{\phi}_k|^2)\mathrm{tr}\,(A_k)\,\mathrm{tr}\,(\rho_k). \tag{108}$$

Here, we have defined

$$|\overline{\phi}_k|^2 := \lim_{T \to \infty} \frac{1}{T} \int_0^T |\phi_k(t)|^2 dt = \frac{1}{d_k^2} \sum_{\nu \in W_k} \text{tr}(P_\nu)^2. \tag{109}$$

Then

$$\mathbb{E}\left(\langle A \rangle_{\rho_U(t)}\right) - \mathbb{E}\left(\langle A \rangle_{\rho_\infty^{UHU^\dagger}}\right) = \sum_{k \neq j} \phi_k(t) \phi_j^*(t) \text{tr}(A_{j,k} \rho_{k,j}) + \\ \sum_k \frac{1}{d_k + 1} \left[\text{tr}(A_k \rho_k) \left(d_k^2 \frac{|\phi_k(t)|^2 - |\overline{\phi}_k|^2}{d_k - 1}\right) + \text{tr}(\rho_k)\text{tr}(A_k)\left(d_k \frac{|\overline{\phi}_k|^2 - |\phi_k(t)|^2}{d_k - 1}\right)\right]. \tag{110}$$

Simplifying this expression, we get

$$\mathbb{E}\left(\langle A \rangle_{\rho_U(t)}\right) - \mathbb{E}\left(\langle A \rangle_{\rho_\infty^{UHU^\dagger}}\right) = \sum_{k \neq j} \phi_k(t) \phi_j^*(t) \text{tr}(A_{j,k} \rho_{k,j}) + \sum_k \frac{d_k}{d_k + 1} F_k(t) \left(\text{tr}(A_k \rho_k) - \frac{\text{tr}(A_k)}{d_k}\text{tr}(\rho_k)\right) \tag{111}$$

where we have defined

$$F_k(t) := d_k \frac{|\phi_k(t)|^2 - |\overline{\phi}_k|^2}{d_k - 1}. \tag{112}$$

This gives

$$R(t) \leq \|A\|_\infty K^2 \max_{k,j} \phi_k(t) \phi_j^*(t) + \max_k F_k(t) \sum_k \left|\text{tr}(A_k \rho_k) - \frac{\text{tr}(A_k)}{d_k}\text{tr}(\rho_k)\right| \tag{113}$$

where we have used

$$\text{tr}(A_{j,k} \rho_{k,j}) = \text{tr}(A \Pi_k \rho \Pi_j) \leq \|A\|_\infty \|\Pi_k \rho \Pi_j\|_1 \leq \|A\|_\infty. \tag{114}$$

By the same reasoning, we can conclude that

$$\text{tr}(A_k \rho_k) \leq \|A\|_\infty \|\rho_k\|_1 \tag{115}$$

and furthermore $\text{tr}(A_k)/d_k \leq \|A\|_\infty$. This allows us to bound

$$\sum_k \left|\text{tr}(A_k \rho_k) - \frac{\text{tr}(A_k)}{d_k}\text{tr}(\rho_k)\right| \leq 2\|A\|_\infty. \tag{116}$$

Altogether, we can conclude that

$$R(t) \leq \|A\|_\infty \left(K^2 \max_{k,j} |\phi_k(t)\phi_j^*(t)| + 2\max_k F_k(t)\right). \tag{117}$$

To make more progress, we need to be able to compute and discuss the functions $\phi_k$. Mathematically speaking, we make a mild spectral assumption expected to be valid in natural quantum many-body systems. We assume that the above characteristic functions decay appropriately.

**Assumption 1** (Spectral assumption). *For some $r > 0$ and $t \leq \text{poly}(N)$, we have*

$$|\phi_k(t)| \leq \frac{1}{t^r}. \tag{118}$$

In this case,

$$R(t) \leq O\left(\|A\|_\infty \frac{K^2}{t^{2r}}\right) \leq O\left(\|A\|_\infty \frac{N^2}{\delta^2 t^{2r}}\right). \tag{119}$$

where we use $K \leq O(N/\delta)$. Such a behavior can be derived, for example, by assuming a physically plausible slowly varying density of states inside each energy window $I_k$: consider the function $E_k(x) : [0,1] \to \mathbb{R}$ defined such that for $x \in \left[\frac{\nu}{d_k}, \frac{\nu+1}{d_k}\right)$, $E_k(x) = E_\nu$. Then, we have

$$\phi_k(t) = \int_0^1 \mathrm{d}x e^{itE_k(x)}. \tag{120}$$

Now, define a linear interpolation of $E_k$: for $x \in \left[\frac{\nu}{d_k}, \frac{\nu+1}{d_k}\right)$

$$\tilde{E}_k(x) = d_k \Delta_\nu x + (\nu+1)E_\nu - \nu E_{\nu+1}, \tag{121}$$

where $\Delta_\nu = E_{\nu+1} - E_\nu$. $\tilde{E}_k(x)$ is piece-wise linear and satisfies $|\tilde{E}_k(x) - E_k(x)| \leq \max_{\nu \in W_k} \Delta_\nu$. In particular, defining

$$\tilde{\phi}_k(t) = \int_0^1 \mathrm{d}x e^{it\tilde{E}_k(x)}, \tag{122}$$

we have $|\phi_k(t) - \tilde{\phi}_k(t)| \leq \max_{\nu \in W_k} \Delta_\nu$. Notice that $\tilde{E}_k(x)$ is strictly increasing hence invertible, and its inverse is continuous and differentiable almost everywhere. By the change of variable $y = \tilde{E}_k(x)$ we get

$$\tilde{\phi}_k(t) = \int_{e_k}^{e_{k+1}} \mathrm{d}y e^{ity} \rho_k(y), \tag{123}$$

where $\rho_k(y) := \frac{\mathrm{d}}{\mathrm{d}y} \tilde{E}_k^{-1}(y)$ is the density of states, and satisfies $\rho_k(y) = \frac{1}{d_k \Delta_\nu}$ for $y \in [E_\nu, E_{\nu+1}]$. We take once again a piecewise linear interpolation of $\rho_k(y)$: for $y \in [E_\nu, E_{\nu+1}]$

$$\tilde{\rho}_k(y) = \frac{y - E_\nu}{d_k \Delta_\nu} \left(\frac{1}{\Delta_{\nu+1}} - \frac{1}{\Delta_\nu}\right) + \frac{1}{d_k \Delta_\nu} \tag{124}$$

is piecewise linear and satisfies

$$|\rho_k(y) - \tilde{\rho}_k(y)| \leq \max_{\nu \in W_k} \frac{1}{d_k} \left|\frac{1}{\Delta_{\nu+1}} - \frac{1}{\Delta_\nu}\right|. \tag{125}$$

Hence, defining

$$\tilde{\tilde{\phi}}_k(t) = \int_{e_k}^{e_{k+1}} \mathrm{d}y e^{ity} \tilde{\rho}_k(y), \tag{126}$$

we have $|\tilde{\tilde{\phi}}_k(t) - \tilde{\phi}_k(t)| \leq \max_{\nu \in W_k} \frac{1}{d_k} \left|\frac{1}{\Delta_{\nu+1}} - \frac{1}{\Delta_\nu}\right|$, and furthermore, integrating by parts gives

$$\tilde{\tilde{\phi}}_k(t) = \frac{1}{it} \int_{e_k}^{e_{k+1}} \mathrm{d}y \tilde{\rho}_k(y) \frac{\mathrm{d}}{\mathrm{d}y} e^{ity} = \frac{1}{it}\left(e^{ite_{k+1}} \tilde{\rho}_k(e_{k+1}) - e^{ite_k}\tilde{\rho}_k(e_k)\right) + \frac{1}{it}\int_{e_k}^{e_{k+1}} \mathrm{d}y \tilde{\rho}'_k(y) e^{ity}. \tag{127}$$

Then

$$\begin{aligned}|\tilde{\tilde{\phi}}_k(t)| &\leq \frac{1}{t}\left|\left(e^{ite_{k+1}} \tilde{\rho}_k(e_{k+1}) - e^{ite_k}\tilde{\rho}_k(e_k)\right)\right| + \frac{1}{t}\left|\int_{e_k}^{e_{k+1}} \mathrm{d}y \tilde{\rho}'_k(y) e^{ity}\right| \\ &\leq \frac{1}{t}\left(2 + \max_{\nu \in W_k} \frac{1}{d_k}\left|\frac{1}{\Delta_{\nu+1}} - \frac{1}{\Delta_\nu}\right|\right).\end{aligned} \tag{128}$$

Overall, we get

$$|\phi_k(t)| \leq \frac{1}{t} + O\left(\max_{\nu \in W_k} \Delta_\nu\right) + O\left(\max_{\nu \in W_k} \frac{1}{d_k}\left|\frac{1}{\Delta_{\nu+1}} - \frac{1}{\Delta_\nu}\right|\right). \tag{129}$$

This shows that our assumption holds with $r = 1$ on time scales such that $1/t$ is large with respect to the spectral gaps and the difference of nearby spectral gaps. This is also compatible with natural expectations of spectral gaps following a Wigner-Dyson distribution, which due to level repulsion favors an almost uniform distribution of the spectral gaps. Altogether, using $K \leq O(N/\delta)$, this proves the following:

**Theorem 7.** *Under Assumption 1, with probability at least $\frac{C}{\epsilon^2} \sum_k \mathrm{tr}\left(\rho_k\right) \frac{1}{d_k}$ for a constant $C$,*

$$\left|\langle A \rangle_{\rho_U(t)} - \langle A \rangle_{\rho_\infty^{UHU^\dagger}}\right| \leq \epsilon + O\left(\|A\|_\infty \frac{N^2}{\delta^2 t^{2r}}\right). \tag{130}$$

### A. Second moments bounds: proofs of Lemmas 5 and 6

We now prove the second moment bounds:

*Proof of Lemma 5.* We have

$$\langle A \rangle_{\rho_U(t)} = \sum_{k,j} \mathrm{tr}\left( A_{j,k} U_k e^{-iH_k t} U_k^\dagger \rho_{k,j} U_j e^{iH_j t} U_j^\dagger \right). \tag{131}$$

Then, let

$$a_{k,j}(U) := \mathrm{tr}\left( A_{j,k} U_k e^{-iHt} U_k^\dagger \rho_{k,j} U_j e^{iHt} U_j^\dagger \right). \tag{132}$$

We have

$$\mathbb{E}\left( \left| \langle A \rangle_{\rho_U(t)} - \mathbb{E}\left( \langle A \rangle_{\rho_U(t)} \right) \right|^2 \right) = \sum_{k,j,r,s} \mathbb{E}(a_{k,j}(U)\bar{a}_{rs}(U)) - \mathbb{E}(a_{k,j}(U))\mathbb{E}(\bar{a}_{rs}(U)) = \sum_{k,j,r,s} \mathrm{Cov}(a_{k,j}(U), a_{rs}(U)). \tag{133}$$

By the Cauchy-Schwarz inequality

$$\mathrm{Cov}(a_{k,j}(U), a_{rs}(U)) \le \sqrt{\mathrm{Var}(a_{k,j}(U))\mathrm{Var}(a_{rs}(U))}, \tag{134}$$

hence

$$\sum_{k,j,r,s} \mathrm{Cov}(a_{k,j}(U), a_{rs}(U)) \le \left( \sum_{k,j} \sqrt{\mathrm{Var}(a_{k,j}(U))} \right)^2 \le K^2 \sum_{k,j} \mathrm{Var}(a_{k,j}(U)) \tag{135}$$

where $K = \sum_k = \Theta(N/\delta)$ is the total number of windows. In the case $k = j$, we have

$$\mathrm{Var}(a_{k,j}(U)) = \mathbb{E}(|a_{kk}(U)|^2) - |\mathbb{E}(a_{kk}(U))|^2 \le \|A\|_\infty \mathrm{tr}\left(\rho_k^2\right) O\left(\frac{1}{d_k}\right) \tag{136}$$

(see ref. [7], Eq. (11)). If $k \ne j$, one has

$$\mathrm{Var}(a_{k,j}(U)) = \mathbb{E}(|a_{k,j}(U)|^2) - |\mathbb{E}(a_{k,j}(U))|^2. \tag{137}$$

A simple computation reveals

$$\mathbb{E}(a_{k,j}(U)) = \mathrm{tr}\left( A \Pi_k \rho \Pi_j \right) \phi_k(t) \bar{\phi}_j(t). \tag{138}$$

The second moment of $a_{k,j}(U)$ can be rewritten as

$$\mathbb{E}(|a_{k,j}(U)|^2) = \mathrm{tr}\left( A_{j,k} \otimes A_{k,j} U_k \otimes U_j e^{-iHt} \otimes e^{iHt} U_k^\dagger \otimes U_j^\dagger \rho_{k,j} \otimes \rho_{j,k} U_j \otimes U_k e^{iHt} \otimes e^{-iHt} U_j^\dagger \otimes U_k^\dagger \right). \tag{139}$$

Another tedious but straightforward computation using Weingarten calculus as outlined in Section II reveals (we omit the time-dependence notation on the $\phi_k(t)$)

$$\begin{aligned}
\mathbb{E}(|a_{k,j}(U)|^2) &= \frac{|\mathrm{tr}\left(A_{j,k}\rho_{k,j}\right)|^2}{(d_k-1)^2(d_j-1)^2}\left[|\phi_k|^2|\phi_j|^2 d_k^2 d_j^2 - d_k^2|\phi_k|^2 - d_j^2|\phi_j|^2 + 1\right] \\
&+ \frac{\|A_{j,k}\rho_{k,j}\|_F^2}{(d_k-1)^2(d_j-1)^2}\left[d_k(1-|\phi_k|^2)(d_j^2|\phi_j|^2-1) + d_j(1-|\phi_j|^2)(d_k^2|\phi_k|^2-1)\right] \\
&+ \frac{\|A_{j,k}\|_F^2 \|\rho_{k,j}\|_F^2 d_k d_j}{(d_k-1)^2(d_j-1)^2}(1-|\phi_k|^2)(1-|\phi_j|^2) \\
&= |\mathrm{tr}\left(A_{j,k}\rho_{k,j}\right)|^2 \left[|\phi_k|^2|\phi_j^2| + O\left(\frac{1}{d_k^2}\right) + O\left(\frac{1}{d_j^2}\right)\right] \\
&+ \|A_{j,k}\rho_{k,j}\|_F^2 \left[O\left(\frac{1}{d_k}\right) + O\left(\frac{1}{d_j}\right)\right] + \|A_{j,k}\|_F^2\|\rho_{k,j}\|_F^2 O\left(\frac{1}{d_k d_j}\right),
\end{aligned} \tag{140}$$

where $\|\cdot\|_F$ indicates the Frobenius norm. Hence

$$\mathbb{E}(|a_{k,j}(U)|^2) - |\mathbb{E}(|a_{k,j}(U))|^2 = |\text{tr}\,(A_{j,k}\rho_{k,j})\,|^2\left[O\left(\frac{1}{d_k^2}\right) + O\left(\frac{1}{d_j^2}\right)\right] \\ + \|A_{j,k}\rho_{k,j}\|_F^2\left[O\left(\frac{1}{d_k}\right) + O\left(\frac{1}{d_j}\right)\right] + \|A_{j,k}\|_F^2\|\rho_{k,j}\|_F^2 O\left(\frac{1}{d_k d_j}\right). \tag{141}$$

Using repeatedly that for matrices $A$ and $B$, $\|AB\|_F \leq \|A\|_\infty \|B\|_F$, we have

$$\|A_{j,k}\rho_{k,j}\|_F^2 = \|\Pi_j A \Pi_k \rho \Pi_j\|_F^2 \leq \|A\|_\infty^2 \|\rho_{k,j}\|_F^2. \tag{142}$$

Furthermore,

$$\|A_{j,k}\|_F^2 = \|\Pi_j A \Pi_k\|_F^2 \leq \min(d_j, d_k)\|A\|_\infty^2, \tag{143}$$

where we have used that $\|\Pi_j\|_F^2 = d_j$ and similarly for $\Pi_k$. Finally, by the Cauchy-Schwarz inequality

$$|\text{tr}\,(A_{j,k}\rho_{k,j})\,|^2 \leq \|A_{j,k}\|_F^2 \|\rho_{k,j}\|_F^2 \leq \min(d_j, d_k)\|A\|_\infty^2\|\rho_{k,j}\|_F^2. \tag{144}$$

Putting everything together we get

$$\text{Var}(a_{k,j}(U)) = \mathbb{E}(|a_{k,j}(U)|^2) - |\mathbb{E}(|a_{k,j}(U))|^2 = \|A\|_\infty^2 \|\rho_{j,k}\|_F^2\left[O\left(\frac{1}{d_k}\right) + O\left(\frac{1}{d_j}\right)\right]. \tag{145}$$

By submultiplicativity of the Frobenius norm, we have that

$$\|\rho_{j,k}\|_F^2 = \|\Pi_j \rho \Pi_k\|_F^2 = \|\Pi_j \rho^{1/2}\rho^{1/2}\Pi_k\|_F^2 \leq \|\Pi_j \rho^{1/2}\|_F^2 \|\rho^{1/2}\Pi_k\|_F^2 = \text{tr}\,(\rho_k)\,\text{tr}\,(\rho_j)\,. \tag{146}$$

Going back to Eq. (135), we have that in both the $k = j$ and $k \neq j$ case, using $\text{tr}\,(\rho_k^2) \leq \text{tr}\,(\rho_k)^2$ for the former case,

$$\text{Var}(a_{k,j}(U)) \leq K^2 \|A\|_\infty^2 \text{tr}\,(\rho_k)\,\text{tr}\,(\rho_j)\left[O\left(\frac{1}{d_k}\right) + O\left(\frac{1}{d_j}\right)\right], \tag{147}$$

which yields

$$\mathbb{E}\left(\left|\langle A\rangle_{\rho_U(t)} - \mathbb{E}\left(\langle A\rangle_{\rho_U(t)}\right)\right|^2\right) \leq \|A\|_\infty^2 O\left(K^2 \sum_k \text{tr}\,(\rho_k)\frac{1}{d_k}\right). \tag{148}$$

$\square$

*Proof of Lemma 6.* We have

$$\langle A\rangle_{\rho_\infty^{UHU^\dagger}} = \sum_k \sum_{\nu \in W_k} \text{tr}\left(P_\nu U_k \rho_k U_k^\dagger P_\nu U_k A_k U_k^\dagger\right). \tag{149}$$

Define $b_k(U)$ as

$$b_k(U) := \sum_{\nu \in W_k} \text{tr}\left(P_\nu U_k \rho_k U_k^\dagger P_\nu U_k A_k U_k^\dagger\right) \tag{150}$$

so that

$$\langle A\rangle_{\rho_\infty^{UHU^\dagger}} = \sum_k b_k(U). \tag{151}$$

The $b_k(U)$ are independent as each only depends on the block $U_k$, hence we have

$$\mathbb{E}\left(\left|\langle A\rangle_{\rho_\infty^{UHU^\dagger}} - \mathbb{E}\left(\langle A\rangle_{\rho_\infty^{UHU^\dagger}}\right)\right|^2\right) = \sum_k \mathbb{E}(b_k(U)^2) - \mathbb{E}(b_k(U))^2. \tag{152}$$

We have

$$\mathbb{E}(b_k(U)) = \frac{|\overline{\phi}_k|^2 d_k^2 - 1}{d_k^2 - 1}\text{tr}\,(A_k\rho_k) + \frac{d_k}{d_k^2 - 1}(1 - |\overline{\phi}_k|^2)\text{tr}\,(A_k)\,\text{tr}\,(\rho_k)$$
$$= \frac{d_k}{d_k^2 - 1}(1 - |\overline{\phi}_k|^2)\text{tr}\,(A_k)\,\text{tr}\,(\rho_k) + \|A\|_\infty \text{tr}\,(\rho_k)\,O\left(\frac{1}{d_k}\right), \quad (153)$$

where we have defined

$$|\overline{\phi}_k|^2 := \lim_{T\to\infty} \frac{1}{T}\int_0^T |\phi_k(t)|^2 \mathrm{d}t = \frac{1}{d_k^2}\sum_{\nu\in W_k}\text{tr}\,(P_\nu)^2 = O\left(\frac{1}{d_k}\right). \quad (154)$$

We move on to the second moment: using the Weingarten formula (see Section II)

$$\mathbb{E}(b_k(U)^2) = \sum_{\nu,\mu\in W_k}\mathbb{E}\left(\text{tr}\left(P_\nu^{\otimes 2}\otimes P_\mu^{\otimes 2} U^{\otimes 4}\rho_k\otimes A_k\otimes \rho_k\otimes A_k U^{\dagger\otimes 4}R_{(1,2)}R_{(3,4)}\right)\right)$$
$$= \sum_{\mu,\nu}\sum_{\sigma,\tau\in S_4}\text{Wg}(\sigma\tau^{-1},d_k)\text{tr}\left(P_\nu^{\otimes 2}\otimes P_\mu^{\otimes 2} R_\sigma R_{(1,2)}R_{(3,4)}\right)\text{tr}\left(R_\tau \rho_k\otimes A_k\otimes \rho_k\otimes A_k\right). \quad (155)$$

To simplify matters, from now on we will assume that the eigenvalues are non-degenerate, that is, $D_\nu = \text{tr}\,(P_\nu) = 1$. As long as $D_\nu$ is at most polynomially large, generalization to the degenerate case is easy but tedious. We denote by $S_2 \times S_2$ the set of permutations in $S_4$ consisting of $(), (1,2), (3,4), ((1,2),(3,4))$, that is, the permutation group of the first two elements and of the second two elements taken separately. Then we have

$$\text{tr}\left(P_\nu^{\otimes 2}\otimes P_\mu^{\otimes 2} R_\sigma R_{(1,2)}R_{(3,4)}\right) = \begin{cases} 1 & \text{if } \sigma \in S_2, \times S_2 \\ \delta_{\mu,\nu} & \text{otherwise.} \end{cases} \quad (156)$$

Then

$$\mathbb{E}(b_k(U)^2) = \sum_{\tau\in S_4}\text{tr}\,(R_\tau\rho_k\otimes A_k\otimes \rho_k\otimes A_k)\left[d_k^2\sum_{\sigma\in S_2\otimes S_2}\text{Wg}(\sigma\tau^{-1},d_k) + d_k\sum_{\sigma\notin S_2\otimes S_2}\text{Wg}(\sigma\tau^{-1},d_k)\right]. \quad (157)$$

This sum contains many terms, but we are only interested in showing that the potentially $O(1)$ terms are compensated by the $O(1)$ term in $\mathbb{E}(b_k(U))^2$ (see Eq. (153)), and that all other terms decay at least as $1/d_k$. We have that $\text{Wg}(\sigma\tau^{-1},d_k) = O(d_k^{-4})$ if $\sigma = \tau$, and $\text{Wg}(\sigma\tau^{-1},d_k) = O(d_k^{-5})$ otherwise (see Table II). This implies that

$$d_k^2\sum_{\sigma\in S_2\otimes S_2}\text{Wg}(\sigma\tau^{-1},d_k) + d_k\sum_{\sigma\notin S_2\otimes S_2}\text{Wg}(\sigma\tau^{-1},d_k) \leq \begin{cases} O(d_k^{-3}) & \text{if } \tau \notin S_2\times S_2, \\ O(d_k^{-2}) & \text{if } \tau \in S_2\times S_2. \end{cases} \quad (158)$$

Together with the fact that

$$\text{tr}\,(R_\tau\rho_k\otimes A_k\otimes \rho_k\otimes A_k) \leq O(d_k^2)\|A\|_\infty^2 \text{tr}\,(\rho_k)^2 \quad (159)$$

we get

$$\sum_{\tau\notin S_2\times S_2}\text{tr}\,(R_\tau\rho_k\otimes A_k\otimes \rho_k\otimes A_k)\left[d_k^2\sum_{\sigma\in S_2\otimes S_2}\text{Wg}(\sigma\tau^{-1},d_k) + d_k\sum_{\sigma\notin S_2\otimes S_2}\text{Wg}(\sigma\tau^{-1},d_k)\right] \leq \|A\|_\infty^2 \text{tr}\,(\rho_k)^2 O(1/d_k). \quad (160)$$

Now consider the permutations $\tau \in S_2\times S_2$. Unless $\tau = ()$ or $\tau = (1,3)$,

$$\text{tr}\,(R_\tau\rho_k\otimes A_k\otimes \rho_k\otimes A_k) \leq O(d_k)\|A\|_\infty^2 \text{tr}\,(\rho_k)^2. \quad (161)$$

Since $(1,3) \notin S_2\times S_2$, this yields

$$\sum_{\tau\neq ()}\text{tr}\,(R_\tau\rho_k\otimes A_k\otimes \rho_k\otimes A_k)\left[d_k^2\sum_{\sigma\in S_2\otimes S_2}\text{Wg}(\sigma\tau^{-1},d_k) + d_k\sum_{\sigma\notin S_2\otimes S_2}\text{Wg}(\sigma\tau^{-1},d_k)\right] \leq \|A\|_\infty^2 \text{tr}\,(\rho_k)^2 O(1/d_k). \quad (162)$$

The only remaining term is the one with $\tau = ()$. We simply bound

$$d_k^2 \sum_{\sigma \in S_2 \otimes S_2} \mathrm{Wg}(\sigma, d_k) + d_k \sum_{\sigma \notin S_2 \otimes S_2} \mathrm{Wg}(\sigma, d_k) \leq d_k^2 \sum_{\sigma \in S_4} \mathrm{Wg}(\sigma, d_k) = \frac{d_k^2}{d_k(d_k+1)(d_k+2)(d_k+3)} \leq \frac{1}{(d_k+1)^2}. \quad (163)$$

Finally,

$$\mathbb{E}(b_k(U)^2) \leq \|A\|_\infty^2 \mathrm{tr}(\rho_k)^2 O\left(\frac{1}{d_k}\right) + \mathrm{tr}(A_k)^2 \mathrm{tr}(\rho_k)^2 \frac{1}{(d_k+1)^2}. \quad (164)$$

In the case of non-degenerate eigenvalues, Eq. (153) reduces to

$$\mathbb{E}(b_k(U)) = \frac{1}{d_k+1} \mathrm{tr}(A_k)\mathrm{tr}(\rho_k) + \|A\|_\infty \mathrm{tr}(\rho_k) O\left(\frac{1}{d_k}\right) \quad (165)$$

which shows

$$\mathbb{E}(b_k(U)^2) - \mathbb{E}(b_k(U))^2 \leq \|A\|_\infty^2 \mathrm{tr}(\rho_k)^2 O\left(\frac{1}{d_k}\right). \quad (166)$$

Then

$$\mathbb{E}\left(\left|\langle A\rangle_{\rho_\infty^{UHU^\dagger}} - \mathbb{E}\left(\langle A\rangle_{\rho_\infty^{UHU^\dagger}}\right)\right|^2\right) \leq \|A\|_\infty^2 \sum_k \mathrm{tr}(\rho_k)^2 O\left(\frac{1}{d_k}\right) \leq \|A\|_\infty^2 O\left(\sum_k \mathrm{tr}(\rho_k) \frac{1}{d_k}\right). \quad (167)$$

□

## V. TRANSLATIONAL INVARIANCE

We will now turn to discuss notions of translational invariance. More generally, we will show that if the original Hamiltonian has a certain symmetry $T : [H, T] = 0$, then the equilibrium state $\rho_\infty^{UHU^\dagger}$ with high probability has the same symmetry in an approximate form. Applying this to the translation symmetry yields the claim in the main text. We will first prove the following. Since we will rely on Lemma 6, we assume non-degenerate eigenvalues, with again the caveat that we expect weak degeneracies not to make a meaningful difference.

**Lemma 8** (Auxiliary lemma). *Let $A$ be an observable such that $\mathrm{tr}(\Pi_i A) = 0$ for all energy windows $I_i$. Let $\epsilon > 0$. Then with probability at least*

$$1 - \frac{\|A\|_\infty^2}{\epsilon^2} \sum_k \Omega\left(\frac{1}{d_k}\right) \mathrm{tr}(\rho_k), \quad (168)$$

*for $U$ drawn from $\mathcal{E}(\delta)$ we have*

$$\left|\mathrm{tr}\left(\rho_\infty^{UHU^\dagger} A\right)\right| \leq \epsilon. \quad (169)$$

*Proof.* We have

$$\Pr\left(\mathrm{tr}\left(\rho_\infty^{UHU^\dagger} A\right)^2 \geq \epsilon^2\right) \leq \frac{1}{\epsilon^2} \mathbb{E}\left(\mathrm{tr}\left(\rho_\infty^{UHU^\dagger} A\right)^2\right). \quad (170)$$

To upper bound $\mathbb{E}\left(\mathrm{tr}\left(\rho_\infty^{UHU^\dagger} A\right)^2\right)$, we proceed exactly as in the proof of Lemma 6. There, we have shown that (see Eq. (164))

$$\mathbb{E}\left(\mathrm{tr}\left(\rho_\infty^{UHU^\dagger} A\right)^2\right) = \sum_k \mathbb{E}(b_k(U)^2) \quad (171)$$

with

$$\mathbb{E}(b_k(U)^2) \leq \|A\|_\infty^2 \mathrm{tr}\,(\rho_k)^2 \, O\left(\frac{1}{d_k}\right) + \mathrm{tr}\,(A_k)^2 \, \mathrm{tr}\,(\rho_k)^2 \, \frac{1}{(d_k+1)^2}. \tag{172}$$

By assumption, in this case $\mathrm{tr}\,(A_k) = 0$, hence

$$\mathbb{E}\left(\mathrm{tr}\left(\rho_\infty^{UHU^\dagger} A\right)^2\right) \leq \|A\|_\infty^2 \sum_k O\left(\frac{1}{d_k}\right) \mathrm{tr}\,(\rho_k) \tag{173}$$

which implies the result. $\square$

As commented earlier, $d_k$ might be smaller at the edge of the spectrum, but it is still expected to be polynomial in the system size, and the tails at the edge of the spectrum may be cut if the state has exponential decay of correlations. Then, we find the following corollary.

**Corollary 2** (Preservation of symmetries). *Suppose the Hamiltonian has no degenerate eigenvalues and let $T$ be a unitary which commutes with the Hamiltonian. Let $A$ be an observable, let $\epsilon > 0$, then with probability at least*

$$1 - \frac{\|A\|_\infty^2}{\epsilon^2} \sum_k \Omega\left(\frac{1}{d_k}\right) \mathrm{tr}\,(\rho_k) \tag{174}$$

*for $U$ drawn from $\mathcal{E}(\delta)$ we have that*

$$|\mathrm{tr}\left((A - TAT^\dagger)\rho_\infty^{UHU^\dagger}\right)| \leq \epsilon. \tag{175}$$

*Proof.* Since $H$ and $T$ commute, $T$ commutes with the projectors onto the eigenspaces of $H$, and in particular with the projectors $\Pi_i$. We then have

$$\mathrm{tr}\left(\Pi_i TAT^\dagger\right) = \mathrm{tr}\left(\Pi_i A\right). \tag{176}$$

We conclude by applying Lemma 8 to the observable $TAT^\dagger - A$. $\square$

As a corollary, we get the statement in eq. (17) in the main text. Let $A$ be a local observable and call $A_C$ the observable $A$ acting on the hypercube $C \in \mathcal{C}_l$. We have for any $C' \in \mathcal{C}_l$

$$\left|\mathrm{tr}\left(A_{C'}(g_\beta(H) - \rho_\infty^{UHU^\dagger})\right)\right| \leq \frac{1}{|\mathcal{C}_l|} \sum_{C \in \mathcal{C}_l} \left|\mathrm{tr}\left(A_C(g_\beta(H) - \rho_\infty^{UHU^\dagger})\right)\right| + \left|\mathrm{tr}\left((A_{C'} - A_C)(g_\beta(H) - \rho_\infty^{UHU^\dagger})\right)\right|$$
$$\leq D_l(\rho_\infty^{UHU^\dagger}, g_\beta(H)) + \frac{1}{|\mathcal{C}_l|} \sum_{C \in \mathcal{C}_l} \left|\mathrm{tr}\left((A_{C'} - A_C)\rho_\infty^{UHU^\dagger}\right)\right|. \tag{177}$$

where we have used that $g_\beta(H)$ is translation invariant. Since $\mathrm{tr}\left((A'_C - A_C)\Pi_k\right) = 0$ for all $k$, as $H$ is translation invariant, by Lemma 8 we have that with high probability

$$\left|\mathrm{tr}\left((A_{C'} - A_C)\rho_\infty^{UHU^\dagger}\right)\right| \leq \epsilon. \tag{178}$$

To see that the probability is exponentially close to 1, assuming all windows in a sub-extensive neighborhood of $\mathrm{tr}\,(\rho H)$ contain exponentially many eigenvalues, we have

$$\frac{\|A\|_\infty^2}{\epsilon^2} \sum_k \Omega\left(\frac{1}{d_k}\right) \mathrm{tr}\,(\rho_k) \leq e^{-\Omega(N)}. \tag{179}$$

---


}

 \end{document}